\documentclass[12pt]{article}

\usepackage{graphicx}
\usepackage{amssymb}

\input epsf

\newcommand{\xx}{\mbox{\boldmath$x$}}

\newcommand{\Bm}{\mbox{\scriptsize B}}

\newcommand{\ppm}{\mbox{\scriptsize p}}
\newcommand{\ttau}{{\boldmath \mbox{$\tau$}}}

\begin{document}

\title{\textbf{Legendre Integrators, Post-Processing and Quasiequilibrium}}
\author{Alexander N. Gorban$^{1,2}$\thanks{agorban$@$mat.ethz.ch, $^{**}$pavelgorban@yandex.ru, $^{***}$ikarlin$@$mat.ethz.ch}, \and  Pavel A. Gorban$^{1,3**}\!\!$, \and and Iliya V. Karlin$^{1,2***}\!\!$,
\\ $^{1}$ ETH-Zentrum, Department of Materials, Institute of Polymers, \\ Sonneggstr. 3, ML J19,
CH-8092 Z{\"u}rich, Switzerland; \\ $^{2}$ Institute of Computational Modeling SB RAS, \\
Akademgorodok, Krasnoyarsk 660036, Russia; \\ $^{3}$ Omsk State University, Omsk, Russia}

\date{}

\maketitle

\begin{abstract}
A toolbox for the development and reduction of the dynamical models of nonequilibrium systems is
presented. The main components of this toolbox are: Legendre integrators, dynamical postprocessing,
and thermodynamic projector.

Thermodynamic projector is the tool to transform almost arbitrary anzatz to a thermodynamically
consistent model, the postprocessing is the cheapest way to improve the solution, obtained by the
Legendre integrators. Legendre Integrators give the opportunity to solve linear equations instead of
nonlinear ones for quasiequilibrium (MaxEnt) approximations.

The essentially new element of this toolbox, the method of thermodynamic projector, is demonstrated
on application to FENE-P model of polymer kinetic theory. The multy-peak model of polymer dynamics is
developed. The simplest example, discussed in details, is the two peaks model for Gaussian manifold
instability in polymer dynamics. This type of models opens a way to create the computational models
for the ``molecular individualism".
\end{abstract}

\clearpage

\tableofcontents

\clearpage

\section {\bf Introduction}

There are many attempts to fill the gap between the microscopic and the macroscopic models (the
famous micro-macro gap),  and to construct closed macroscopic equations. Most of the closure
assumptions have a relatively narrow domain of applicability, and their usage has the following
problems:

1) Violation of the basic physics (thermodynamics) laws;

2) Absence of the accuracy control procedures;

3) Absence of the successive step-by-step procedure of the refinement of a model.

The main object of investigation is the evolution equation
\begin{eqnarray}\label{eqn1}
\dot{\Psi}=J(\Psi);
\end{eqnarray}
where $J$ is some operator, and $\Psi$ is the distribution function over the phase space.

The constructed methods are aimed at extracting the dynamics of the macroscopic variables from the
microscopic equations (\ref{eqn1}). The prototypes of these methods are the quasiequilibrium
approximation, dual integrators and the thermodynamic projector.

The quasiequilibrium closure for the set of the macroscopic variables $M(\Psi)$ is built with the
help of the solution to the variation problem (MaxEnt approximation)\footnote{From time to time it is
discussed in the literature, who was the first to introduce the quasiequilibrium approximations, and
how to interpret them. At least a part of the discussion is due to a different r\^{o}le the
quasiequilibrium plays in the entropy--conserving and the dissipative dynamics. The very first use of
the entropy maximization dates back to the classical work of G.\ W.\ Gibbs \cite{Gibb}, but it was
first claimed for a principle by E.\ T.\ Jaynes  \cite{Janes1}. Probably the first explicit and
systematic use of quasiequilibria to derive dissipation from entropy--conserving systems is due to
the works of D.\ N.\ Zubarev. Recent detailed exposition is given in \cite{Zubarev}. For dissipative
systems, the use of the quasiequilibrium to reduce description can be traced to the works of H.\ Grad
on the Boltzmann equation \cite{Grad}. The viewpoint of two of the present authors (ANG and IVK) was
influenced by the papers by L.\ I.\ Rozonoer and co-workers, in particular, \cite{KoRoz,Ko,Roz}. A
detailed exposition of the quasiequilibrium approximation for Markov chains is given in the book
\cite{G1} (Chapter 3, {\it Quasiequilibrium and entropy maximum}, pp.\ 92-122), and for the BBGKY
hierarchy in the paper \cite{Kark}. We have applied maximum entropy principle to the description the
universal dependence the 3-particle distribution function $F_3$ on the 2-particle distribution
function $F_2$ in classical systems with pair interactions \cite{BGKTMF}. A very general discussion
of the maximum entropy principle with applications to dissipative kinetics is given in the review
\cite{Bal}.  The methods for corrections the quasiequilibrium approximations are developed in
\cite{GK1,GKTTSP94,KTGOePhA2003,Plenka}.}:
\begin{eqnarray}\label{eqn2}
S(\Psi)\rightarrow max\nonumber\\
\\
M(\Psi)=M,\nonumber
\end{eqnarray}
where $S(\Psi)$ is the entropy.

The quasiequilibrium closure is always thermodynamically consistent, but the problem 2 (the absence
of the accuracy control) remains unsolved, and the problem 3 (the absence of the refinement
procedures) can be solved by adding new macroscopic variables to the problem (\ref{eqn2}). But
uncontrolled enlargement of the macroscopic variables set give us no guarantee of the accuracy
improvement. There exists one more specific problem for the quasiequilibrium approximation
(\ref{eqn2}). Usually while solving the variation problem (\ref{eqn2}) we can find explicit
dependencies $\Psi(\Lambda)$ and $M(\Lambda)$, where $\Lambda$ are the corresponding Lagrange
multipliers (dual variables), more or less easily. Much more difficult is to find the dependencies
$\Lambda(M)$ and $\Psi(M)$ which we need for the closure of the macroscopic equations.

The method of the Legendre integrators consists of building and solving the equations of motion for
the dual variables. The methods of the first order, based on this idea were suggested and tested in
the papers \cite{IKOePhA02, IKOePhA03, GKIOeNONNEWT2001}

The method of the thermodynamic projector let us to represent every ansatz-manifold as the solution
to the variation problem (\ref{eqn2}) with the specially chosen constraints.

The thermodynamic projector is the unique operator which transforms the arbitrary vector field
equipped with the given Lyapunov function into a vector field with the same Lyapunov function (and
also this happens on any manifold which is not tangent to the level of the Lyapunov function).

Equations which are derived by the method of the thermodynamic projector are always {\bf
thermodynamically consistent}. Although this idea was published in the year 1992 \cite{GK1}, the full
construction is published only recently in application to the chemical kinetics \cite{InChLANL}.

One of the problems, discussed in this paper, is to construct the method of the thermodynamic
projector for the derivation of the physically consistent macroscopic equations for the polymer
dynamics. In the process of building the thermodynamic projector and the quasiequilibrium
approximation is involved the Lyapunov function for the equations (\ref{eqn1}) which is the entropy
$S$. The equations for the polymer dynamics (Fokker-Planck equation) allows us to use the huge amount
of different Lyapunov functions and each of them can be formally chosen to describe the macroscopic
processes. We need to analyze the different Lyapunov functions for the Fokker-Planck equation.

The problem of the accuracy estimation of the resulting approximations and their further improvement
is suggested to solve with the procedures of the post-processing.

Suppose that for the dynamical system (\ref{eqn1}) the approximate invariant manifold has been
constructed and the approximate slow motion equations $\Psi_{M}(t)$ have been derived:

\begin{equation}\label{slag}
\frac{d\Psi_{M}}{dt} = P_{\Psi_{M}}(J(\Psi_{M})),
\end{equation}

\noindent where $P_{\Psi_{M}}$ is the corresponding projector onto the tangent space $T_{\Psi_{M}}$
of $\Psi_{M}$. Suppose that we have solved the system (\ref{slag}) and have obtained $\Psi_{M}(t)$.
Let's consider the following two questions:

\begin{itemize}
\item{How well this solution approximates the true solution $\Psi(t)$ given the same initial conditions?}
\item{How is it possible to use the solution $\Psi_{M}(t)$ for it's refinement without solving the system (\ref{slag}) again?}
\end{itemize}

These two questions are interconnected. The first question states the problem of the  accuracy
estimation. The second one states the problem of post-processing.

The corresponding methods to answer these questions are developed and described in this work.

{\section {\bf Elimination of fast variables with the help of the Lyapunov function}}

The most popular way to investigate the dynamics of complicated systems is to split the motion into
the slow and the fast components, and then to exclude the fast component. As a result, one gets a
system of equations that describes the evolution of the slow variables. The necessary conditions of
usefulness of this method are usually formulated as a set of restrictions for the possible dynamics
of the ``fast subsystem". Here the ``fast subsystem" is the subsystem which describes the evolution
of the fast variables with an assumption that slow variables are constant.

Unfortunately, often appear situations where we cannot avoid using this method, and there is no proof
that it is valid. These situations appear almost everywhere in physical kinetics. Here one follows
with the same scheme: the relaxation processes are splitted into slow and fast. In spite of the fact
that in most cases the proofs of validity of this scheme are absent, the experience helps to avoid
fatal errors.

In this section the method to obtain the equations of the macrokinetics from the microdescription is
demonstrated. The basis of the analysis is the assumption that if the macroscopic variables are
chosen in the proper way, then all other variables relax fast: the probability distribution of the
microscopic variables after a small period of time is determined with good accuracy by the
macroscopic variables. Let us call this assumption the ``quasiequilibrium hypothesis".

The notion ``macroscopic variables" is a somewhat relative and is introduced to stress the difference
of these variables from ``everything else". For example, one-particle distribution function can be
``macroscopic" for the full description of the system.

The goal of this section is to describe the most primitive procedure of derivation of the equations
for the slow variables and to discuss the form of these equations.

In this paper the reduction of description goes on with the help of the Lyapunov functions. This
formalism is the case of the known principle of the conditional maximum of entropy with given values
of the macroscopic variables.

Let us review the basic notions of the convex analysis which are used here.

The subset $U$ of the vector space $E$ is convex, if for every two points $x_{1},x_{2}\in U$ it
contains the segment between $x_{1}$ and $x_{2}$: for every $\lambda\in[0,1]$
\begin{equation}\label{oprvypmn}
\lambda x_{1}+(1-\lambda)x_{2}\in U.
\end{equation}
The intersection of any number of the convex sets is convex.

The convex envelope of the subset $M$ of a vector space $E$ is the smallest convex set $co M\subset
E$, that includes $M$. It is the intersection of all the convex sets, that include $M$.

If the set $U\subset E$ is convex , $x_{1},...,x_{k}\in U$, $\lambda_{1},...,\lambda_{k}\geq0$,
$\sum_{i}\lambda_{i}=1$, then $\sum_{i}\lambda_{i}x_{i}\in U$. It leads to another definition of the
convex envelope:
\begin{equation}\label{oprconenv}
coM=\{\sum_{i=1}^{k}\lambda_{i}x_{i}|x_{1},...,x_{k}\in M,\:\:
\lambda_{1},...,\lambda_{k}\geq0,\:\:\sum_{i}\lambda_{i}=1, k<\infty\}.
\end{equation}
If $dimE=n$, then in the equation (\ref{oprconenv}) it is sufficient to take $k\leq n+1$ (Carthedory
theorem).

The function $f$, defined on the convex set $U\subset E$, is convex, if its epigraph, i.e. the set of
pairs
\begin{equation}\label{oprnadgr}
Epif=\{(x,g)|x\in U, g\geq f(x)\},
\end{equation}
is the convex set in $E\times R$. Sometimes it is convenient to consider functions which can reach
the value $f=\infty$. If there occurs a necessity to study the functions $f$ which are defined on the
non-convex set $V\subset E$, then it is supposed that $f$ is convex, if the restriction of $f$ onto
every convex subset of $V$ is convex. If the restriction of $f$ onto every line segment from the
region of definition is convex, then $f$ is convex. The differentiable function $f$ of the class
$C^{2}$ is convex if and only if the matrix of the second derivatives $\partial^{2}f/\partial
x_{i}\partial x_{j}$ is nonnegative defined (i.e. all its eigenvalues are nonnegative). The smooth
convex function $f$ on the convex set $U\subset R^{n}$ satisfies the inequality
\begin{equation}\label{neqconv}
f(x^{1})-f(x^{2})\geq(\nabla f|_{x^{2}},x^{1}-x^{2})=\sum_{i}(\partial f/\partial
x_{i})_{x=x^{2}}(x^{1}_{i}-x^{2}_{i}), (x^{1},x^{2}\in U).
\end{equation}
Geometrically it means, that the graph of $f$ is located above the hyperplane, tangent at the point
$x=x^{2}$.

The function $f$ is called strictly convex if in the domain of the definition there is no line
segment, on which it is constant and finite ($f(x)=const\neq\infty$). The sufficient condition for
the differentiable function $f$ of the $C^{2}$ class to be strictly convex is that the matrix of the
second derivatives $\partial^{2}f/\partial x_{i}\partial x_{j}$ is positive defined.

In the set of the maximum points of the convex function $f$ on the compact set $U$ ($U$ may be not
convex) there are some boundary points of $U$, and if $U$ is convex, then there are some extreme
points of $U$. The set of the minimum points of $f$ on the convex set $U$ is convex (but may be
empty). The strictly convex continuous function has its maximum only in the boundary points of $U$,
and if $U$ is convex, then in the extreme points. The strictly convex function may have the finite
minimum only in one point.

The function $f$ called concave if the function $-f$ is convex.

Every bounded convex function on the open subset of $R^{n}$ is continuous.

Let the $C^{2}$-smooth function $H$ be defined in the domain $U\subset R^{n}$. Let us correspond the
vector $\mu=\nabla_{x}H:\mu_{i}=\partial H/\partial x_{i}$ to every point $x\in U$. If the matrix
$\partial\mu_{i}/\partial x_{j}=\partial^{2}H/\partial x_{i}\partial x_{j}$ is non-degenerated, then
for the transform $x\rightarrow\mu$ there locally (in the neighborhood of every point) exist the
differentiable inverse transform. The variables $\mu$ are often called conjugated variables, and the
transform $x\rightarrow\mu$ is called ``transition to the conjugated coordinates". Let the transform
$x\rightarrow\mu$ be invertible on the open set $V\subset U$. This means that the function $x(\mu)$
is defined on $V$. Assuming the smoothness of this function, we describe the inverse transform
$\mu\rightarrow x$ in the same way as the direct. For this purpose we introduce a function
\begin{eqnarray}\label{legtrans}
G(\mu)=(\mu,x(\mu))-H(x(\mu))=\sum_{i}\mu_{i}x_{i}(\mu)-H(x(\mu)), \nonumber \\ \frac{\partial
G}{\partial\mu_{i}}=x_{i}+\sum_{j}\mu_{j}\frac{\partial
x_{j}}{\partial\mu_{i}}-\sum_{j}\frac{\partial H}{\partial x_{j}}\frac{\partial
x_{j}}{\partial\mu_{i}}=x_{i}.
\end{eqnarray}
The function $G$ called the Legendre transform of $H$.

With the help of the conjugated coordinates it is possible to write down the necessary conditions of
the extremum for the problems with the linear constraints on the open set in a very simple way:
\begin{equation}\label{mtsk}
\begin{array}{l}
H(x)\rightarrow min,\\ \sum_{j}m_{ij}x_{j}=M_{i},\: (i=1,...,k),\: x\in U.
\end{array}
\end{equation}
With the method of Lagrange multipliers we get the system of the equations which is giving us the
necessary conditions for the solution to the problem (\ref{mtsk}):
\begin{equation}\label{nmcond}
\begin{array}{l}
\mu_{j}=\sum_{i}\lambda_{i}m_{ij}, \: j=1,...,n,\\  \sum_{j}m_{ij}x_{j}=M_{i}, \: (i=1,...,k),
\end{array}
\end{equation}
where the $\lambda_{i}$ are the Lagrange multipliers. The necessary conditions for the extremum are
given by the system of the equations (\ref{nmcond}). One part of the system is linear in the $x$
coordinates, and the other part is linear in the conjugated coordinates $\mu$.

Let us have the Legendre transform $G(\mu)$ for the function $H(x)$, let the transform
$x\rightarrow\mu$ have the smooth reverse, and let the solution to the problem (\ref{mtsk}) be unique
for some open set of the values of the vector $(M_{1},...,M_{k})\in R^{n}$. Also let the point of the
minimum $x_{min}$, and, consequently, the minimal value of $H$ be smooth dependent on $M$,
$H_{min}=H(M)$. Let us denote $\mu_{M_{i}}=\partial H(M)/\partial M_{i}$,
$\mu_{M}=(\mu_{M_{1}},...,\mu_{M_{k}})$. Let us get some information about the function $H(M)$ from
the functions $H(x)$ and $G(\mu)$ without solving any equations. With the known value of the vector
$\mu_{M}$ we can immediately find the vector $\mu$ in the corresponding point of the conditional
minimum, $\mu_{j}=\sum_{i}\mu_{M{i}}m_{ij}$. From this equality we get
\begin{equation}\label{xotmum}
x(\mu_{M})=(\nabla_{\mu}G(\mu))|_{\mu_{j}=\sum_{i}\mu_{M_{i}}m_{ij}}.
\end{equation}
From $x_{\mu_{M}}$ we obtain $M(\mu_{M})$ and $H(M(\mu_{M}))$:
\begin{equation}\label{himotxmum}
M_{i}(\mu_{M})=\sum_{j}m_{ij}x_{j}(\mu_{M}), H(M(\mu_{M}))=H(x(\mu_{M})).
\end{equation}
Finally, the Legendre transform $G(\mu_{M})$ for the function $H(M)$ is:
\begin{equation}\label{ltrotxmu}
G(\mu_{M})=(\mu_{M},M(\mu_{M}))-H(M(\mu_{M}))=G(\mu(\mu_{M})).
\end{equation}
So, we can find dependencies $\mu(\mu_{M})$, $x(\mu_{M})$, $M(\mu_{M})$, $H(\mu_{M})$ and
$G(\mu_{M})$ from the functions $H(x)$ and $G(x)$ without solving any equations. We hope, that the
similar notations for $H(x)$ and corresponding conditional minimum function $H(M)$, and for their
Legendre transforms $G(\mu)$ and $G(\mu_{M})$ will not cause a confusion. Let us note, that with our
assumptions the reversibility of the transform $M\rightarrow\mu_{M}$ follows from the reversibility
of the transform $x\rightarrow\mu$, and moreover, the function $M(\mu_{M})$ can be found explicitly.

The convexity of the function $H(x)$ usually makes our assumptions (existence and uniqueness of the
conditional minimum, global reversibility of the transform $x\rightarrow\mu$, smoothness of the
function $H(M)$) easier to check. Note, that the convexity of the function $H(M)$ is neither
necessary nor sufficient condition for our assumptions. If $H(x)$ is convex, then the function of the
conditional minimum $H(M)$ is convex too.

Now we proceed to the problem of elimination of the fast variables. Let us have the system of
differential equations with smooth right-hand sides
\begin{equation}\label{sisdif}
\dot{x}=F(x),
\end{equation}
in the convex domain $U\subset R^{n}$, and moreover let the linear transform $x\rightarrow M$,
$M_{i}=\sum_{j}m_{ij}x_{j}$ from the phase space to the space of the slow variables $M$ be defined.
We can assume, that we have no linearly dependent rows in the matrix $m_{ij}$, because it is always
possible to eliminate the linear dependent functions $M_{i}(x)$, if they are present.

Let us assume that in the interesting for us domain of the initial conditions $x_{0}$ the solutions
$x(t)$ of the equations (\ref{sisdif}) are developing in the following way: the vector $x(t)$ is
going fast to the value which is defined by the slow variables $M$; after that $x$ can be represented
as the function of $M$ with a good accuracy, and this function is unique for every initial
conditions. So,

A)For each value of the slow variables $M\in M(U)$ there exist such $x=x^{*}(M)$, that if
$M(x^{0})=M^{0}$, then $x(t)$ is going very fast to some small neighborhood of the $x^{*}(M^{0})$, and
during that $M(x(t))$ is almost constant.

B)In the process of the further evolution, $x(t)$ stays in the small neighborhood of the value of $x$
which corresponds to $M(x(t))$, so $x$ is close to $x^{*}(M(x(t)))$.

It is usually impossible to give a strong proof for A and B for the situations of real complexity in
the nonequilibrium thermodynamics, so this assumptions are, probably, the weakest point of all the
construction. We are accepting them because we are sure that the evolution of the macroscopic
variables is possible to describe by the autonomous system of differential equations of the first
order (if it is impossible, then, probably, one should extend the list of the macroscopic variables
with respect to the physical properties of the investigated process). There is another way to deal
with this problem: to equip the obtained approximations by the {\it postprocessing}. The
postprocessing helps us to correct the errors, if they are not too big, and gives us a signal if they
are too big.

If we know the function $x^{*}(M)$, then we can write
\begin{equation}\label{equaproc}
\dot{M}=mF(x^{*}(M)),\:\: \dot{M}_{i}=\sum_{j}m_{ij}F_{j}(x^{*}(M)).
\end{equation}
In general, this equation can be used only for short periods of time which do not exceed some limit.
The right-hand side $mF(x^{*}(M))$ of the equations (\ref{equaproc}) is not exactly the $mF(x(t))$,
and it may cause the error increment, and as a result the solution of the equations (\ref{equaproc})
will divert from the true solution strongly. The exclusion is the case when in accordance to the
equations (\ref{equaproc}) $M(t)$ tends to the only stable fixed point when $t\rightarrow\infty$. If
the solution of the equations (\ref{equaproc}) and the real values of $M(x(t))$ are not succeed in
getting far one from another during the time in which the solution of the equations (\ref{equaproc})
is coming in the small neighborhood of the fixed point, then the equations (\ref{equaproc}) can be
used also for $t\rightarrow\infty$.

The function $x^{*}(M)$ for the particular system is not unique, but the range of choice is small in
that sense, in which the neighborhood of $x^*(M(x(t)))$ (in which the evolution goes after the short
period of time) is small.

Let us have the Lyapunov function $H(x)$ for the system (\ref{sisdif}) which is decreasing along the
trajectories. We can try to find the dependence $x^{*}(M)$ as the solution to the problem
$H(x)\rightarrow min$, $mx=M$. This way seems to be natural, but it does not follow directly from the
assumptions A and B. For example, there could be a situation in which $H$ is very sensitive to small
changes of the slow variables, and not sensitive to the changes of the fast variables. In this
situation the assumption, that $x^{*}(M)$ is the point of conditional minimum of the function $H$,
may not give the desired result. The following idea does not solve the problem, but it can be useful:
In the applications, the system (\ref{sisdif}) usually dependens on some parameters. It seems to be
more reasonable to use the Lyapunov function which does not depend on these parameters, if there
exists such a function. It is most important in the case when among the parameters we have such, that
their values are determining, whether is it possible to split the variables to fast and slow, or not.

So, the fast variables will be eliminated with the help of the Lyapunov function. Let us have the
Lyapunov function $H$ for the initial system, let the transform $x\rightarrow\mu=\nabla_{x}H$ have
the smooth inverse, and let us know the Legendre transform $G(\mu)$ for the function $H(x)$. Here it
is also assumed that for every $M\in M(U)$ the problem (\ref{mtsk}) has the unique solution, and the
minimum point $x^{*}(M)$, and the function of the conditional minimum $H(M)$  smoothly depend on $M$.
With the value $\mu_{M}=\nabla_{M}H(M)$ it is possible to find $\mu(\mu_{M})$, $x(\mu(\mu_{M}))$
(look at the (\ref{xotmum}-\ref{ltrotxmu})). The result is
\begin{equation}\label{mtchkn}
\dot{M}=mF(\nabla_{\mu}G(\mu))|_{\mu=\mu_{M}m},
\end{equation}
where $\mu_{M}m$ is the product of the row vector $\mu_{M}$ and the matrix $m$:
\begin{eqnarray*}
(\mu_{M}m)_{j}=\sum_{i}\mu_{M_{i}}m_{ij},
\end{eqnarray*}
$\nabla_{\mu}G$ is the vector with the components $\partial G/\partial\mu_{i}$, and all derivatives
are taken in the point $\mu=\mu_{M}m$. The right-hand sides of (\ref{mtchkn}) are defined as the
functions of $\mu_{M}$. In order to define them as functions of $M$, one needs to make the Legendre
transform, find the function $H(M)$ and, respectively, $\mu_{M}=\nabla_{M}H(M)$ from the function
$G(\mu_{M})$ (\ref{ltrotxmu}). It is impossible to make these calculations  explicitly in such a
general case. It seems to be a very natural and convenient to define the right-hand sides of the
kinetic equation as the functions of the conjugated variables. If in the beginning the right-hand
sides of the equation (\ref{sisdif}) are defined as the functions of $\mu$  (i.e. $\dot{x}=J(\mu)$),
then the equations (\ref{mtchkn}) have a very simple form:
\begin{equation}\label{mtchkn1}
\dot{M}=mJ(\mu_{M}m).
\end{equation}
$H(M)$ is the Lyapunov function for (\ref{mtchkn}), its time derivative due to the system
(\ref{mtchkn}) is not positive:
\begin{equation}\label{teors}
\dot{H}(M)=(\mu_{M}, mJ(\mu_{M}m))=(\mu_{M}m, J(\mu_{M}m))\leq0,
\end{equation}
because $(\mu,J(\mu))=\dot{H}(x)\leq0$.

Let us call the systems dissipative, if $\dot{H}\leq0$  and conservative, if $\dot{H}=0$. For the
dissipative system we have $\dot{H}(M)\leq0$ (\ref{teors}), and if the system is conservative, then
for all values of $\mu$ we have $(\mu,J(\mu))=\dot{H}(x)=0$, and then from the equation (\ref{teors})
we get $\dot{H}(M)=(\mu_{M}, mJ(\mu_{M}m))=(\mu_{M}m, J(\mu_{M}m))=0$. So, we proved the following

{\bf Theorem}\footnote{This is a rather old theorem, one of us had published this theorem in 1984
already as textbook material (\cite{G1}, chapter 3 ``Quasiequilibrium and entropy maximum", p. 37,
see also the paper \cite{GKIOeNONNEWT2001}), but from time to time different particular cases of this
theorem are continued to be published as new results.}. {\it The Lyapunov function for the
microscopic system remains the Lyapunov function for the macroscopic system, and if the microscopic
system is conservative, then its projection to the space of the macroscopic variables remains
conservative.}

If necessary, it is easy to perform further exclusion of the variables in the equations
(\ref{mtchkn}) with the help of the function $H(M)$. The right-hand sides of the resulting equations
will be defined again as the functions on the conjugated variables, and the function of the
conditional minimum will be the Lyapunov function again. Let us note that in (\ref{mtchkn1}) we have
neither $H$ nor $G$ in the explicit form (they occur only when we need to find the connections
between $M$ and $\mu_{M}$ or $x$ and $\mu$).

Convexity of $H$ was never used above, but the natural domain of applicability of the described
formalism are systems with convex Lyapunov functions $H(x)$, or at least with such $H$, that the sets
$\{x|H(x)<h\}$ are convex. Otherwise there exist such linear manifolds, that the local minimum of $H$
is not unique on them, and further considerations are required to select the relevant minima. The
finite dimensionality of the phase space is not so important, because everything said above can be
applied to the infinite-dimension case with proper restrictions. Let $E$ be the Banach space,
$U\subset E$ be the convex open set, $H:U\rightarrow R$ be $C^{2}$-smooth function. With every point
$x\in U$ we associate the linear functional $\mu_{x}\in E^{*}$: $\mu_{x}=\nabla_{x}H$ which is the
differential of $H$ in the point $x$. Let $V$ be the set of the values of $\mu_{x}$ for $x\in U$ and
let us have the smooth mapping $J$ from $E^{*}$ to $E$ in the neighborhood of $V$. The system
$(U,H,J)$ determines the system of equations:
\begin{equation}\label{sdinf}
\dot{x}=J(\mu_{x}).
\end{equation}
Let $L$ be the closed subset of $E$ and for every $M\in U/L$ let the problem $H(x)\rightarrow min,\:
x/L=M,\: x\in U$ have the unique solution $x_{min}$ which is $C^{2}$-smooth dependent on $M$,
$H(M)=H(x_{min})$. Denoting $\mu_{M}=\nabla_{M}H(M)\in(E/L)^{*}\in E^{*}$ we can define the
factor-system which is the exact analogue of (\ref{mtchkn}):
\begin{equation}\label{faksis}
\dot{M}=J(\mu_{M})/L.
\end{equation}
Here the argument $J$ is the linear functional on $\mu_{M}$: $\mu_{M}x=\mu_{M}(x/L)$

The described procedure of the elimination of variables has one very important commutativity
property: If one makes a further simplification and transact to the variables $N=N(M)$, then after
the application of the described formalism to the system (\ref{faksis}) with the function $H(M)$, one
get the same result as after the application of this formalism directly to the reduction from $x$ to
$N(x)=N(M(x))$. So, the chain of exclusions $x\rightarrow M\rightarrow N$ gives us the same result as
the direct exclusion $x\rightarrow N$.

\section{\bf The main problems in usage of the quasiequilibrium approximations}

Our problem is to build the closed system
\begin{eqnarray*}
\dot{M}=J(M),
\end{eqnarray*}
from the initial system (\ref{eqn1}) and its Lyapunov function.

If we know the function $x^{*}(M)$ then it is sufficient to calculate $m(F(x^{*}(M)))$. This problem
is the problem of the calculation of the projection of the microscopic vector field $F$ on the
macroscopic variables $M$ in known point $x^{*}(M)$. Let us call this problem the problem about the
macroscopic projection. If the right-hand parts are expressed through $\mu$ then we have the problem
about the macroscopic projection too.

Another problem is to find $\mu_{M}$. Usually it is necessary to solve the system of non-linear
equations (if the function $H$ is not quadratic) to solve this problem. Indeed, let us consider the
conditions for the conditional extremum of $H$ with given values of the moments $M$. From the
functions $H(x)$, $G(\mu)$ we get $\mu(\mu_{M})$, $x(\mu_{M})$, $M(\mu_{M})$, $H(M(\mu_{M}))$. But in
this list we have no function $\mu_{M}(M)$. We can find this function as the solution of the equation
\begin{equation}\label{eq16}
M(\mu_{M}m)=M.
\end{equation}

Let us give a few examples.

{\it One-particle approximation}. Let $x$ be the $N$-particle distribution function,
$f_{N}(\xi_{1},...,\xi_{N})$, where $\xi_{i}$ is vector of coordinates and momenta of the $i$-th
particle, and let the evolution of this function be described by the linear equation
\begin{equation}\label{eq17}
\frac{\partial f_{N}}{\partial t}=Lf.
\end{equation}
Furthermore, let $M$ be one-particle distribution function
\begin{equation}
f_{1}(\xi)=N\int f_{N}(\xi,\xi_{2},...,\xi_{N})d\xi_{2}...d\xi_{N},
\end{equation}
and $H$ be the entropy (we use the $H$-function which is equal to the minus entropy)
\begin{equation}\label{eq18}
H(f_{N})=\int f_{N}(\ln f_{N}-1)d^{N}\xi,
\end{equation}
For given $f_{N}$, $H$, $f$ we get $\mu=\ln f_{N}$, $f_{N}=\exp{\mu}$,
\begin{equation}\label{eq19}
G(\mu)=\int\exp{\mu(\xi_{1},...,\xi_{N})}d^{N}\xi,
\end{equation}
$m(f_{N})=\int\sum_{i=1}^{N}\delta(\xi-\xi_{i})f_{N}(\xi_{1},...,\xi_{N})d^{N}\xi$; the extremum
conditions (\ref{nmcond}) are of the form
\begin{eqnarray}\label{eq20}
\mu(\xi_{1},...,\xi_{N})=\int d\xi
\mu_{1}(\xi)\sum_{i=1}^{N}\delta(\xi-\xi_{i})=\sum_{i}\mu_{1}(\xi_{i}), \nonumber \\
f_{N}=\exp{\sum_{i}\mu_{1}(\xi_{i})}.
\end{eqnarray}

The normalization condition here is $\int f_{N}d^{N}\xi=1$, that is
\begin{equation}
\int\exp{\mu_{1}(\xi)}d\xi=1.
\end{equation}
Connection between the macroscopic variables $f_{1}$ (that is $M$) and the quasiequilibrium values of
the microscopic variables $f^{*}_{N}$ (that is, $x^{*}_{M}$) is given by well known formula:
\begin{equation}\label{eq21}
f_{N}(\xi_{1},...,\xi_{N})=\frac{1}{N^{N}}f_{1}(\xi_{1})...f_{1}(\xi_{N}).
\end{equation}
Projection of the microscopic vector field (\ref{eq17}) can be found by direct integration.

{\it Two-particle distribution function as the macroscopic variable}. One-particle distribution
function $f_{1}(\xi)$ is often not sufficient because, for example, the energy of the interaction of
pairs of particles cannot be found from this function. Much more detailed description is given by the
two-particle distribution function
\begin{equation}\label{eq22}
f_{2}(\xi_{1},\xi_{2})=N(N-1)\int f_{N}(\xi_{1},...,\xi_{N})d\xi_{3}...d\xi{N}.
\end{equation}
We can easily find the expression
\begin{eqnarray*}
\mu(\xi_{1},...,\xi_{N})=\sum_{i,j,i\neq j}\mu_{2}(\xi_{i},\xi_{j});\\
f_{N}(\xi_{1},...,\xi_{N})=\exp{\mu}=\exp{\sum_{i,j,i\neq j}\mu_{2}(\xi_{i},\xi_{j})}.
\end{eqnarray*}
But it is difficult to find the connection between $\mu_{2}$ and $f_{2}$ explicitly. Only a series
expansion  for it in the neighborhood of the uncorrelated state is known \cite{BGKTMF}. The problem
about the macroscopic projection becomes hard too: the necessary integrals in general case are
impossible to find analytically. For two-particle distribution functions as well as for majority of
the most interesting variables the transform $M\leftrightarrow\mu_{M}$ is very complicated in the
direct direction and not very simple (as simple as the derivation of $f_{2}$ from $f_{N}$) in the
opposite direction.

So, we need to avoid the necessity to calculate $\mu_{M}(M)$ (and, if possible, to make less
calculations to find $M(\mu_{M})$).

The first of these two problems (avoiding calculation of $\mu_{M}(M)$) is solved by the method of the
Legendre integrators which is developed by us \cite{IKOePhA02}.

\section{\bf Legendre integrators}

The main idea of the Legendre integrators is to find some alternate way to solve the macroscopic
equations $\dot{M}=J(x)$: a way to find their solution in the absence of the explicit form of these
equations.

First of all, note, that we have a linear connection between $\dot{M}$ and $\dot{\mu}_{M}$:
\begin{equation}\label{conn}
\frac{dM}{dt}=(m(D^{2}_{x}S(x))^{-1}m^{T})\frac{d\mu_{M}}{dt};
\end{equation}
\begin{eqnarray}\label{mjumu}
\dot{M}=\frac{d}{dt}(mx(\mu_{M}m))=m(D_{\mu}x)m^{T}\mu;\nonumber\\\\
D_{\mu}x=(D_{x}\mu)^{-1}=(D^{2}_{x}S(x))^{-1}\nonumber.
\end{eqnarray}

Calculation of the functions $m(F(x))$ is the standard problem of the macroscopic projection.
Dependencies $x(\mu_{M})$ are usually quite simple. We suggest the following advancing in time to
solve (unknown) equations $\dot{M}=\Phi(M)$:
\begin{equation}\label{dual}
\mu_{M}(t)\rightarrow x=x(\mu_{M})\rightarrow
\dot{M}\rightarrow\dot{\mu}_{M}\rightarrow\mu_{M}(t+\Delta t)\rightarrow M(t+\Delta t).
\end{equation}
In the sequence (\ref{dual}) there is one operation of macroscopic projection and one operation of
solving the system of linear equations (\ref{conn}).

Formally, it is possible to write down the equations for $\mu_{M}$:
\begin{equation}\label{mudot}
\frac{d\mu_{M}}{dt}=(m(D^{2}_{x}S(x))^{-1}m^{T})^{-1}mF(x),
\end{equation}
where $x=x^{*}_{M}$.

Nevertheless, explicit inversion of the operator in the right-hand part of the equation (\ref{mudot})
is usually difficult and one should use the chain of computations (\ref{dual}). In our first
calculations using of the Legendre integrators \cite{IKOePhA02,IKOePhA03} the methods of the first
order of accuracy were used. This is not the principal restriction: the scheme (\ref{dual}) gives us
a possibility to calculate $\dot{\mu}_{M}$ for any given $\mu_{M}$, so all known methods of the
higher order can be used (for example, the Runge-Kutta method with the different procedures of the
automatic step selection \cite{Gustafsson, Hairer, Hairer2}).

\section{\bf Lyapunov functions for the Fokker-Planck equation}

The Fokker-Planck equation (FPE) in the absence of the drive forces has the form
\begin{equation}\label{FPE}
\frac{\partial\Psi(q,t)}{\partial t}=\nabla_{q}\{D(\Psi(q,t)\nabla_{q}U(q)+\nabla_{q}\Psi(q,t))\},
\end{equation}
where $\Psi$ is the probability density over the configuration space, $q$ is the point of this space,
$\Psi(q)$ is the function of the time $t$, $U(q)$ is the normalized potential energy
($U=U_{potential}/kT$), $D(q)$ is positively semidefinite diffusion operator ($(y_{i},D_{y})\geq0$).

The FPE has two important properties:

1) Conservation of the total probability:
\begin{equation}\label{procons}
\frac{d}{dt}\int\Psi(q,t)dq\equiv0.
\end{equation}

2) Dissipation: for every convex function of one variable $h(a)$ ($h''(a)>0, a\geq0$) the following
functional $S[\Psi]$ is monotonically non-increasing in time:
\begin{equation}\label{Lyapfunc}
S[\Psi]=-\int\Psi^{*}(q)h\left(\frac{\Psi(q)}{\Psi^{*}(q)}\right)dq,
\end{equation}
where
\begin{equation}\label{equilFPE}
\Psi^{*}(q)=const\cdot\exp(-U(q)),
\end{equation}
is the Boltzmann-Gibbs distribution.

For $h(a)=a\ln a$, the functional $S[\Psi]$ is the usual Boltzmann-Gibbs-Shannon entropy:
\begin{equation}\label{SBGS}
S[\Psi]=-\int\Psi^{*}(q)\ln\left(\frac{\Psi(q)}{\Psi^{*}(q)}\right)dq,
\end{equation}
Let us calculate the time derivative of $S[\Psi]$ due to FPE (\ref{FPE}). Note, that
\begin{eqnarray*}
\nabla_{q}\left(\frac{\Psi(q)}{\Psi^{*}(q)}\right)=\frac{\nabla_{q}\Psi(q)+\Psi(q)\nabla_{q}U}{\Psi^{*}(q)},
\end{eqnarray*}
so we can rewrite FPE as follows:
\begin{eqnarray*}
\frac{\partial\Psi(q,t)}{\partial
t}=\nabla_{q}D\left(\Psi^{*}(q)\nabla_{q}\left(\frac{\Psi(q)}{\Psi^{*}(q)}\right)\right).
\end{eqnarray*}
Let us consider FPE in the domain $\Omega$. Function $dS/dt$ consists of two summands: The first is
the integral of the local ``production of $S$", $\int\sigma(q)dq$, and the second is the flow through
the boundary of the domain $\Omega$:
\begin{eqnarray*}
\frac{dS(\Psi)}{dt}=-\int_{\Omega}h'\left(\frac{\Psi}{\Psi^{*}}\right)\nabla_{q}\left(D\Psi^{*}\left(\nabla_{q}\left(\frac{\Psi}{\Psi^{*}}\right)\right)\right)dq=\\
-\int_{\Omega}div
\left[h'\left(\frac{\Psi}{\Psi^{*}}\right)D\Psi^{*}\nabla_{q}\left(\frac{\Psi}{\Psi^{*}}\right)\right]dq+\int_{\Omega}\Psi^{*}h''\left(\frac{\Psi}{\Psi^{*}}\right)\left(\nabla_{q}\left(\frac{\Psi}{\Psi^{*}}\right),D\nabla_{q}\left(\frac{\Psi}{\Psi^{*}}\right)\right)dq=\\
\int_{\partial\Omega}\Psi^{*}h'\left(\frac{\Psi}{\Psi^{*}}\right)\left(\nu_{q},D\nabla_{q}\left(\frac{\Psi}{\Psi^{*}}\right)\right)dw+\int_{\Omega}\sigma(q)dq,
\end{eqnarray*}
where $dw$ is the differential of the area, $\nu_{q}$ is a vector of the unitary normal to
$\partial\Omega$ in the point $q$, $\sigma(q)$ is the entropy $S$ production:
\begin{equation}\label{sigma}
\sigma(q)=\Psi^{*}h''\left(\frac{\Psi}{\Psi^{*}}\right)\left(\frac{\Psi}{\Psi^{*}},D\nabla_{q}\left(\frac{\Psi}{\Psi^{*}}\right)\right)\geq0.
\end{equation}
Let the flow of $\Psi$ through the boundary $\partial\Omega$ be equal to zero:
\begin{eqnarray*}
\left(\nu_{q},D\nabla_{q}\left(\frac{\Psi}{\Psi^{*}}\right)\right)=0,
\end{eqnarray*}
at all points of $\partial\Omega$. Then
\begin{eqnarray*}
\frac{dS}{dt}=\int_{\Omega}\sigma(q)dq\geq0.
\end{eqnarray*}
The most important cases of $S$ selection are:

$h(q)=a\ln a$, $S$ is the Boltzmann-Shannon-Gibbs entropy;

$h(a)=a\ln ax-\alpha\ln ax$  is the maximal family of {\it additive trace-form} entropies
\cite{ENTR1,ENTR2,ENTR3} (these entropies are additive for composition of independent subsystems);

$h(a)=\frac{1-a^{\alpha}}{1-\alpha}, \alpha\neq1$ is the Tsallis entropy \cite{Abe}. These entropies
are not additive, but become additive after nonlinear monotonous transformation. This property can
serve as definition of the Tsallis entropies in the class of generalized entropies (\ref{Lyapfunc})
\cite{ENTR3}.

\section{\bf Macroscopic variables and quasiequilibrium distribution functions for FPE}

The set of the macroscopic variables can be continuous or discrete. Let $\alpha$ be the discrete or
continuous parameter, that enumerates the macroscopic variables, and $M_{\alpha}$ be the
corresponding variables. Every macroscopic value $M_{\alpha}$ is defined by its microscopic density
$m_{\alpha}(q):$
\begin{equation}\label{macro}
M_{\alpha}=\int_{\Omega}m_{\alpha}(q)\Psi(q)dq
\end{equation}
The choice of the domain $\Omega$, in which we are solving the FPE, needs to be discussed separately.
We can suppose formally, that $\Omega=R^{n}$, but for the calculations it is better to make it as
small as possible with the preservation of the accuracy. Usually, when $\|q\|\rightarrow\infty$ the
function $\Psi(q)$ tends to zero faster, then exponential, and we can a priori select the bounded
domain $\Omega$, out of which $\Psi$ is negligibly small.

We shall do the calculations for the general form of $S$ (see equation (\ref{Lyapfunc})) and give the
examples for the most popular choice (\ref{SBGS}) of $S$.

Quasiequilibrium function $M_{\alpha}$ for the given Lyapunov function $S$ (\ref{Lyapfunc}) is
defined as the solution to the problem
\begin{equation}\label{maxent}
\left\{\begin{array}{lc}S(\Psi)\rightarrow\max&\\ \int
m_{\alpha}(q)\Psi(q)dq=M_{\alpha}&\end{array}\right.,
\end{equation}

Due to the convexity of $h$ (and, consequently, concavity of $S$), it is sufficient to investigate
the conditions of the local extremum:
\begin{equation}\label{extnes}
D_{\Psi}S=\sum_{\alpha}m_{\alpha}(q)\mu_{\alpha},
\end{equation}
where $\mu_{\alpha}$ are variables, dual to $M_{\alpha}$ ($\mu_{M}$). For continuous parameter the
sum in the equation (\ref{extnes}) is replaced by integration on $\alpha$.

Next, we use the standard Riesz representation of functionals (through the $L^{2}$ scalar product).
Let us write
\begin{eqnarray*}
D_{\Psi}S(\Psi)=-h'\left(\frac{\Psi}{\Psi^{*}}\right);\\
h'\left(\frac{\Psi}{\Psi^{*}}\right)=-\sum_{\alpha}m_{\alpha}(q)\mu_{\alpha}.
\end{eqnarray*}
For the quasiequilibrium distribution we have
\begin{equation}\label{QEFPE}
\Psi=\Psi^{*}g\left(-\sum_{\alpha}m_{\alpha}(q)\mu_{\alpha}\right),
\end{equation}
where $g(a)$ is a function of one variable, inverse to $h'(b)$. Note, that $h'(b)$ is a monotonous
increasing function (because $h$ is convex), so $g(a)$ is a monotonous increasing function too, and
$g'(a)=(h''(g(a)))^{-1}$.

Let us denote the quasiequilibrium distribution function (\ref{QEFPE}) as
$\Psi^{qe}(\{\mu_{\alpha}\},q)$.

For the BGS entropy $h(b)=b(\ln b-1)$, $h'(b)=\ln b$, $g(a)=\exp a$, and the equations (\ref{QEFPE})
transforms into the following equation:
\begin{equation}\label{QEFPEB}
\Psi^{qe}(\{\mu_{\alpha}\},q)=\Psi^{*}\exp\left(-\sum_{\alpha}m_{\alpha}(q)\mu_{\alpha}\right).
\end{equation}

For the next steps it is convenient to consider the temperature dependence explicitly (i.e. write
$\beta U$ instead of $U$ in FPE, $\beta=1/kT$), then we have $\Psi^{*}=const\cdot\exp(-\beta U)$.

For the classical BGS entropy (\ref{SBGS}) the quasiequilibrium distribution will take the simplest
form:
\begin{equation}\label{psiqe}
\Psi^{qe}(\{\mu_{\alpha}\},q)=\exp\left(-\mu_{0}-\mu_{U}U-\sum_{\alpha}m_{\alpha}(q)\mu_{\alpha}\right),
\end{equation}
where $\mu_{U}=\beta=1/kT$, $\mu_{0}$ is a variable, conjugated to $M_{0}=\int_{\Omega}\Psi
dq\equiv1$. The function (\ref{psiqe}) is a solution to the problem:
\begin{equation}\label{Clas}
\left\{\begin{array}{l}-\int_{\Omega}\Psi\ln\Psi dq\rightarrow\max\\
M_{0}(\Psi)=\int_{\Omega}\Psi(q)dq\nonumber=1\\
M_{U}(\Psi)=\int_{\Omega}U(q)\Psi(q)dq\nonumber=M_{U}\\
M_{\alpha}(\Psi)=\int_{\Omega}m_{\alpha}(q)\Psi(q)dq\nonumber=M_{\alpha}\end{array}\right.
\end{equation}
In the problem (\ref{Clas}) we move from the relative (so-called Kullback) entropy to the absolute
entropy.

Selection of the macroscopic variables is the most critical point in construction of the
quasiequilibrium approximations. It is always necessary to select them, basing on the specific of the
problem. Nevertheless, there are some simple general recommendations about construction of the set of
variables for the Legendre integrators.

1) It is necessary to include $M_{0}$ in the list of variables, because $\mu_{0}$ is not constant in
time;

2)It is useful to include $M_{U}$ in the list of variables. With this variable in the process of the
relaxation all other $\mu_{\alpha}\rightarrow0$, and $\mu_{U}\rightarrow1/kT$.

3) It is better for the set of the functions $m_{\alpha}(q)$ to be linearly independent.

For the classical entropy we have
\begin{equation}\label{QEPsi}
\Psi^{qe}(\{\mu\},q,t)=\exp\left(-\mu_{0}(t)-\mu_{U}(t)U(q)-\sum_{\alpha}m_{\alpha}(q)\mu_{\alpha}(t)\right).
\end{equation}
Due to the equation (\ref{QEPsi}) we have
\begin{equation}\label{Psidot1}
\frac{\partial\Psi}{\partial
t}=-\Psi\left[\frac{d\mu_{0}}{dt}+U(q)\frac{dM_{U}}{dt}+\sum_{\alpha}m_{\alpha}(q)\frac{\mu_{\alpha}}{dt}\right];
\end{equation}
The FPE gives us
\begin{eqnarray}\label{Psidot2}
&&\frac{\partial\Psi}{\partial t}=\nabla D\left(\Psi^{*}\nabla\frac{\Psi}{\Psi^{*}}\right)=
-\Psi\left.\mbox{\Huge [}(\mu_{U}-\beta)(\nabla,D\nabla)U(q)+\right. \nonumber\\
&&\sum_{\alpha}\mu_{\alpha}(\nabla,D\nabla)m_{\alpha}(q)-
\sum_{\alpha}(2\mu_{U}\mu_{\alpha}-\beta\mu_{\alpha})(\nabla U(q),D\nabla m_{\alpha}(q))-\\
&&\left.\sum_{\alpha,\alpha'}\mu_{\alpha}\mu_{\alpha'}(\nabla m_{\alpha}(q),D\nabla
m_{\alpha'}(q))\right.\mbox{\Huge ]}\nonumber.
\end{eqnarray}
To calculate $\frac{dM}{dt}(\{\mu\})$ means to calculate the following integrals:
\begin{eqnarray*}
\frac{dM_{U}}{dt}=\int_{\Omega}U(q)\frac{\partial\Psi(q)}{\partial t}dq;\\
\frac{dM_{\alpha}}{dt}\int_{\Omega}m_{\alpha}(q)\frac{\partial\Psi(q)}{\partial t}dq,
\end{eqnarray*}
where $\frac{\partial\Psi}{\partial t}$ is calculated due to equation (\ref{Psidot2}), $dM_{0}/dt=0$.

From the equation (\ref{Psidot1}) we get the conditions for derivation of $\dot{\mu}$
\begin{eqnarray}\label{mudoteq}
-\frac{d \mu_{0}}{d
t}-M_{U}\frac{d\mu_{U}}{dt}-\sum_{\alpha}M_{\alpha}\frac{d\mu_{\alpha}}{dt}=\dot{M}_{0}=0;\nonumber\\
-M_{U}\frac{d\mu_{0}}{dt}-\langle U^{2}\rangle_{\Psi}\frac{d\mu_{U}}{dt}-\sum_{\alpha}\langle
Um_{\alpha}\rangle_{\Psi}=\dot{M}_{U};\\ -M_{\alpha}\frac{d\mu_{0}}{dt}-\langle
Um_{\alpha}\rangle_{\Psi}\frac{d\mu_{U}}{dt}-\sum_{\gamma}\langle
m_{\gamma}m_{\alpha}\rangle_{\Psi}\frac{d\mu_{\gamma}}{dt}=\dot{M}_{\alpha},\nonumber
\end{eqnarray}
where by $\langle f(q)g(q)\rangle_{\Psi}$ we denote the averaging $\langle
fg\rangle_{\Psi}=\int_{\Omega}f(q)g(q)\Psi(q)dq$.

We get the closed system for derivation of the dynamics of $\mu$. But the question about the choice
of the macroscopic variables still remains open.

In the problem of the quasiequilibrium we find the projections of $\Psi$ to the given set of the
functions (linear space), afterwards we calculate $\Psi$ due to the maximum entropy condition.

It seems to be physically sensible to choose the additional variables to $M_{0}, M_{U}$ as {\it the
projections of  $\Psi$ onto some equilibrium states}:
\begin{equation}\label{malpha}
M_{\alpha}(\Psi)=\int_{\Omega}e^{-\alpha U(q)}\Psi(q)dq.
\end{equation}
There are two classical choices of macroscopic variables:

1) $\alpha=R_{+}$ (Laplace transform of the energy distribution density)

2) $\alpha=ik, k\in R$ (Fourier transform of the energy distribution density).

The variable $M_{U}$ is the average energy in the potential well $U(q)$. In analogue to this, the
variable $M_{\alpha}(\Psi)$ (\ref{malpha}) for the real $\alpha>0$ can be considered as the energy in
the potential well $e^{-\alpha U(q)}$. This potential is gained by the monotonous nonlinear
deformation of the energy scale $U\rightarrow e^{-\alpha U(q)}$. For imaginary $\alpha$ this
nonlinear deformation is given by the periodical functions $U\rightarrow \cos(kU)+i\sin(kU)$

A benefit of usage of (\ref{malpha}) is also in that $\langle m_{ \alpha}m_{ \alpha'} \rangle =M_{
\alpha+ \alpha'}$, and we have to perform less calculations in (\ref{mudoteq}). This set of the
deformed energies can be used for both the initial potential $U$ and the set of additional
potentials.

Is this set of macroscopic variables sufficient for description of nonequilibrium kinetics of
polymers in presence of flow? Probability densities for all the quasiequilibrium distributions which
can be constructed with this macroscopic variables have the form $\Psi(q)=\varphi(U(q))$, where
$\varphi(U)$ is a function of one variable. Is this class of distributions sufficient for the
specific problem? This question can be answered only after specification the problem. But what is
possible to do, if the closure with these variables gives too big error (the estimation of accuracy
is discussed below)? There are at least two ways: to extend the list of variables or to improve the
quasiequilibrium manifold \cite{GKIOeNONNEWT2001,GKTTSP94} (application of the methods of invariant
manifolds to improving the quasiequlibrium closure for dynamics of dilute polymeric solution is
presented in \cite{ZKD2000}). The extension of the list of variables is the central method of the
extended irreversible thermodynamics \cite{EIT}. It is possible to combine the potential energy
$U(q)$, the vector of the configuration space $q$, and the gradient of $U(q)$, $\nabla U(q)=- F(q)$,
($F(q)$ is the force) and to obtain a huge amount of densities $m(q)$ which can be scalars, vectors,
or tensors. The corresponding ``macroscopic variables" are $\int_{\Omega}m(q)\Psi(q) dq$.

The best hint for a choice of new macroscopic variables is the analysis of the right hand side of
dynamic equations \cite{GKPRE96}. The well known distinguished macroscopic variable associated with
the polymeric kinetic equations is the polymeric stress tensor \cite{Bird,Martin}. This variable is
not the conserved quantity but nevertheless it should be treated as a relevant slow variable because
it actually contributes to the macroscopic (hydrodynamic) equations. Equations for the stress tensor
are known as the constitutive equations, and the problem of reduced description for the polymeric
models consists in deriving such equations from the kinetic equation.

The tensor
\begin{equation}
\label{tau_p} \ttau_{\ppm \, ij}=k_{\Bm}T \left(\delta_{ij} - \int_{\Omega}F_iq_j\Psi(q) dq\right)
\end{equation}
gives a contribution to stresses caused by the presence of polymer molecules for unit density. Here
$F(q)=-\nabla U(q)$ is the force vector, $\delta_{ij}$ is the Kronecker symbol. For spherically
symmetric potentials ($U(q)=u(q^2)$) this tensor is symmetric. The tensor of dencities
$m_{ij}(q)=F_i(q)q_j$ is the first addition to the dencities which depend only of $U(q)$.

\section{\bf Macroscopic variables and boundary conditions}

There is a  standard technique to solve the boundary value and initial-boundary value problems of
mathematical physics: first to build the space of the functions which satisfy the boundary
conditions, and then to find the solution in this space.

When one uses the Legendre integrators, a special technique is needed to satisfy the boundary
conditions.

FPE describes the evolution of the probability distribution. It conserves the total probability. The
natural boundary conditions for the FPE is the absence of the flow through the boundary of $\Omega$:
\begin{equation}\label{BC}
\Psi^{qe}\left(\nu_q,D\nabla_{q}\left(\frac{\Psi}{\Psi_{q}}\right)\right)=0,
\end{equation}
on $\partial\Omega$, where $\nu_q$ is a vector of outlet normal to $\partial\Omega$ in the point $q$.

Quasiequilibrium distribution functions (\ref{QEPsi}) satisfy the condition (\ref{BC}), if
\begin{equation}\label{BCQE}
\left\{\begin{array}{l} (\nu_q,D\nabla_{q}U(q))=0\\
(\nu_q,D\nabla_{q}m_{\alpha}(q))=0\end{array}\right.,
\end{equation}
for all $\alpha$.

There is also a different way to satisfy conditions (\ref{BC}): to make
$\Psi^{*}|_{\partial\Omega}=0$. It is possible to do by making $U(q)\rightarrow\infty$ while
$q\rightarrow q_{0}\in\partial\Omega$. But this choice leads to the singularities and is very
inconvenient from the numerical point of view.

Conditions (\ref{BCQE}) look somewhat surprisingly, if considered without the context of the
quasiequilibrium approximations: for the quasiequilibrium solutions the absence of the flow through
the barrier follows not from the infinite heights of the barrier, but from the fact, that the normal
derivatives of $U$ and $m_{\alpha}$ are zeros.

To satisfy the condition (\ref{BCQE}) it may be necessary to deform the initial potential $U$ and
densities $m(q)$. This deformation will be the smoothing of $U$ near $\partial\Omega$. The error,
introduced by this deformation is usually not very big (because of the smallness of $\Psi^{*}$ near
$\partial\Omega$) and can be estimated easily.

So, the quasiequilibrium approximation and the Legendre Integrators of any order of accuracy are
built, and the way to satisfy the boundary conditions is suggested. First numerical experiments
\cite{IKOePhA02,IKOePhA03} proved the effectiveness of this idea.

The main computational challenge in this method is to calculate the integrals of the form
\begin{equation}\label{Int}
\int_{\Omega}\left(\sum\lambda_{k}\varphi_{k}(q)\right)\exp\left(\sum\gamma_{i}\psi_{i}(q)\right)dq
\end{equation}
where $\varphi_{k}(q), \psi_{i}(q)$ are known functions (usually they are given analytically). For
the problems of the polymer physics the complexity of the problem (\ref{Int}) is dependent on few
characteristics:

1) The quantity of the different functions $\varphi_{k}(q), \psi_{i}(q)$ is usually 5-10;

2) The dimension of the space in which the integration goes is usually 10-100.

\section{\bf Thermodynamic projector and Galerkin approximations}

Almost every manifold of the functions can be represented as the solution to the quasiequilibrium
problem (\ref{maxent}), if this manifold is not tangent to the level surface of the entropy $S=const$
\cite{GK1}. For this representation only the right system of restrictions is needed. By the simple
parameterization with the moments $M(\Psi)$ it is possible to get only the classical quasiequilibrium
manifolds (\ref{maxent}).  The restrictions which are necessary to represent manifold $\Omega$ as the
quasiequilibrium manifold are built as follows. Let $f\in\Omega$, and $T_{f}$ be the tangent space to
$\Omega$ in the point $f$. On the space of the distribution functions $E$ we define the projector
$P_{f}:E\rightarrow T_{f}$. Operator $P_{f}$ depends smoothly on the point $f$ and on $T_{f}$. The
problem of the quasiequilibrium is posed as follows:
\begin{equation}\label{QEPre}
\left\{\begin{array}{l} S(\Psi)\rightarrow\max\\ P_{f}(\Psi-f)=0\end{array}\right.
\end{equation}
The necessary and sufficient condition for $f$ to be the unique solution to the problem (\ref{QEPre})
is \cite{GK1}:
\begin{equation}\label{CondQE}
\ker P_{f} \subseteq \ker DS|_{f},
\end{equation}
that is, if $P_{f}(\varphi)=0$, then $DS|_{f}(\varphi)=0$. For the classical entropy
$DS|_{f}(\varphi)=-\int\varphi(q)\ln f(q)dq$ and the condition (\ref{CondQE}) takes the form:
\begin{equation}\label{CondQE1}
\mbox{If }P_{f}(\varphi)=0,\mbox{ then }\int\varphi\ln fdq=0.
\end{equation}

Among all projectors which satisfy the condition (\ref{CondQE}) there is unique projector which has
the following property: let us have the appropriate equation
\begin{eqnarray*}
\dot{\Psi}=J(\Psi),
\end{eqnarray*}
for which $dS[\Psi]/dt\geq0$. Then for the projected equation on $\Omega$
\begin{equation}\label{TDPro}
\dot{f}=P_{f}(J(f)),
\end{equation}
we also have $dS[f]/dt\geq0$.

This projector was introduced in the paper \cite{InChLANL}, and there it is also proved its
uniqueness. It is built as follows.

Let us require that the field of projectors, $P(\Psi,T)$, is defined for any $\Psi$ and $T$, if

\begin{equation}\label{transversality}
  T\not{\!\subset} \ker D_{\Psi}S.
\end{equation}

From these conditions it follows immediately that in the equilibrium, $P(\Psi^*,T)$ is the orthogonal
projector onto $T$ (orthogonality with respect to entropic scalar product $\langle |
\rangle_{\Psi^*}$).

The field of projectors is constructed in the neighborhood of the equilibrium based on the
requirement of maximal smoothness of $P$ as a function of $g_{\Psi}=D_{\Psi}S$ and $\Psi$. It turns
out that to the first order in the deviations $\Psi-\Psi^*$ and $g_{\Psi}-g_{\Psi^*}$, the projector
is defined uniquely. Let us first describe the construction of the projector, and next discuss its
uniqueness.

Let the subspace $T\subset E$, the point $\Psi$, and the differential of the entropy in this point,
$g_{\Psi}=D_{\Psi}S$, be defined such that the transversality condition (\ref{transversality}) is
satisfied. Let us define $T_0=T\bigcap\ker g_{\Psi}$. By the condition (\ref{transversality}),
$T_0\neq T$. Let us denote, $e_g=e_g(T)\in T$ the vector in $T$, such that $e_g$ is orthogonal to
$T_0$, and is normalized by the condition $g(e_g)=1$. Vector $e_g$ is defined unambiguously.
Projector $P_{S,\Psi}=P(\Psi,T)$ is defined as follows: For any $z\in E$,

\begin{equation}\label{projgen}
  P_{S,\Psi}(f)=P_0(z)+e_gg_{\Psi}(f),
\end{equation}
where $P_0$ is the orthogonal projector on $T_0$ (orthogonality with respect to the entropic scalar
product $\langle |\rangle_{\Psi}$). Entropic projector (\ref{projgen}) depends on the point $q$
through the $\Psi$-dependence of the scalar product $\langle |\rangle_{\Psi}$, and also through the
differential of $S$ in $\Psi$, the functional $g_{\Psi}$.

Obviously, $P(f)=0$ implies $g(f)=0$, that is, the thermodynamicity requirement is satisfied.
Uniqueness of the thermodynamic projector (\ref{projgen}) is supported by the requirement of the
\textit{maximal smoothness} (analyticity) \cite{InChLANL} of the projector as a function of
$g_{\Psi}$ and $\langle |\rangle_{\Psi}$, and is done in two steps which we sketch here:

\begin{enumerate}
  \item Considering the expansion of the entropy in the
  equilibrium up to the quadratic terms, one shows that in the
  equilibrium the thermodynamic projector is the orthogonal
  projector with respect to the scalar product $\langle|\rangle_{\Psi^*}$.
  \item For a given $g$, one considers auxiliary dissipative
  dynamic systems which satisfy the condition:
  For every $\Psi'\in U$, it holds, $g_{\Psi}(J(\Psi'))=0$, that is,
  $g_{\Psi}$ defines an additional linear conservation law for the
  auxiliary systems. For the auxiliary systems, the point $\Psi$ is
  the equilibrium. Eliminating the linear conservation law $g_{\Psi}$,
  and using the result of the previous point, we end up with the
  formula (\ref{projgen}).
\end{enumerate}

Thermodynamic projector allows to use almost arbitrary manifolds as quasiequilibrium closure
assumption. If the projection of FPE (\ref{TDPro}) is built with the thermodynamic projector, then
$dS/dt$ conserves (not only the sign, but also the value). The only restriction is that the manifold
must not be tangent to the level surfaces of $S$ (and must contain the equilibrium point).

Let us write down the explicit formulas for the closure assumption of the form
\begin{equation}\label{Ga}
f(q)=\Psi^{*}(q)+\sum_{\alpha}f_{\alpha}(q)\mu_{\alpha}.
\end{equation}
Due to probability conservation for all $\alpha$ we have $\int f_{\alpha}(q)dq=0$.

Tangent spaces to the manifold (\ref{Ga}) in all points coincide and have the form
$T=\{\sum_{\alpha}\mu_{\alpha}f_{\alpha}(q)\}$. The natural coordinates in $T$ are $\mu_{\alpha}$.
For every $f(q)$ of the form (\ref{Ga}) there is the entropic scalar product, defined in $T$:
\begin{eqnarray*}
\langle\varphi|\psi\rangle_{f}=-\langle\varphi|(D^{2}S|_{f})\psi\rangle=\int\frac{\varphi(q)\psi(q)}{f(q)}dq
\end{eqnarray*}
In the coordinates $\mu_{\alpha}$ this scalar product has the form
\begin{eqnarray*}
\langle\sum_{\alpha}f_{\alpha}(q)\mu_{\alpha}|\sum_{\beta}f_{\beta}(q)\mu'_{\beta}\rangle_{f}=\sum_{\alpha,\beta}g_{\alpha,\beta}\mu_{\alpha}\mu'_{\beta},
\end{eqnarray*}
where
\begin{eqnarray*}
g_{\alpha,\beta}=\int\frac{f_{\alpha}(q)f_{\beta}(q)}{f(q)}dq.
\end{eqnarray*}

We will need the orthonormalized basis of the subspace $T\bigcap\ker(DS|_{f})$. This subspace is
defined by the equation
\begin{eqnarray*}
\int\sum_{\alpha}f_{\alpha}(q)\mu_{\alpha}\ln\frac{f(q)}{\Psi^{*}(q)}dq=0.
\end{eqnarray*}
Let be $\int f_{1}(q)\ln\frac{f(q)}{\Psi^{*}(q)}dq\neq0$ bor the definiteness. Suppose for $\alpha>1$
\begin{eqnarray}\label{proor}
q_{\alpha}=f_{\alpha}-\nu_{\alpha}f_{1},\\\mbox{where } \nu_{\alpha}=\frac{\int
f_{\alpha}(q)\ln\frac{f(q)}{\Psi^{*}(q)}dq}{\int f_{1}(q)\ln\frac{f(q)}{\Psi^{*}(q)}dq}\nonumber
\end{eqnarray}
Let us orthogonalize the family of the vectors $q_{\alpha}$ ($\alpha>1$) with respect to the scalar
product $\langle\cdot|\cdot\rangle_{f}$. We will get the orthogonal basis in $T\bigcap\ker(DS|_{f})$:
$\{e_{\alpha}\} (\alpha>1)$.

Let $e_{1}\in T$ be the vector, orthogonal to all $e_{\alpha}$ (for example,
$e_{1}=a(f_{1}-\sum_{\alpha>1}e_{\alpha}\langle f_{1}|e_{\alpha}\rangle_{f})$) and let $e_{1}$ be
normalized in the following way: $\int e_{1}(q)\ln\frac{f_{1}(q)}{\Psi^{*}(q)}dq=1$. The projection
of the vector $J$ on $T$ is defined in this way:
\begin{equation}\label{PrTD}
P^{th}_{f}J=e_{1}\int
J(q)\ln\frac{f_{1}(q)}{\Psi^{*}(q)}dq+\sum_{\alpha>1}e_{\alpha}\int\frac{J(q)e_{\alpha}(q)}{f(q)}dq.
\end{equation}

Projector (\ref{PrTD}) allows to consider every manifold of the form (\ref{Ga}) which  is not tangent
to the level surface of the entropy $S$, as the quasiequilibrium manifold. If the vector field is
projected with the operator (\ref{PrTD}), then the dissipation is conserved.

As we can see, there is a ``law of the difficulty conservation": for the quasiequilibrium with the
moment parameterization the manifold is not  explicit, and it can be difficult to calculate it.
Thermodynamic projector completely eliminates this difficulty. From the other side, on the
quasiequilibrium manifold with the moment parameterization (if it is found) it is easy to find the
dynamics: simply write $\dot{M}_{\alpha}=\int\mu_{\alpha}Jdq$. The building of the thermodynamic
projector may require some efforts.

Finally, for each  of the distributions $\Psi$ it is easy to find its projection on the classical
quasiequilibrium manifold $\Psi\rightarrow\Psi^{qe}_{M(\Psi)}$: it requires just calculation of the
moments $M(\Psi)$. The analogue projection for the general thermodynamic projector is rather
difficult: $\Psi\rightarrow f$ with the condition $P^{th}_{f}(\Psi-f)=0$. This equation defines the
projection of some neighborhood of the manifold $\Omega$ on $\Omega$, but the solution of this
equation is rather difficult. Fortunately, we need to build such operators only to analyze the fast
processes of the initial relaxation layer, and it is not necessary to investigate the slow dynamics.

\section{\bf A few words about the specifics of the computational difficulties}

From the computational point of view, the main difficulties in realization of the described methods
are in the calculation of the integrals of the form
\begin{eqnarray*}
\int_{\Omega}\sum a_{i}f_{i}(q)F(\sum b_{j}f_{j}(q))dq
\end{eqnarray*}
where $f_{i}$ are given functions of the vector $q$, $a_{i}, b_{i}$ are the numbers, $F$ is a
function of one variable. The usual $F$ are $F(z)=e^{z}; F(z)=1/z,...$. The usual dimension of
$\Omega$ in polymer physics is a few hundreds, number of different $f_{i}$ is a few dozens.

In any case, the transition from the integration of the whole FPE to solution of the moment equations
gives a considerable decrease of the computation time.

In the methods of Legendre integrators and thermodynamic projector the computational problems of the
linear algebra are present: the solution of the system of linear equations $C\dot{\mu}=\dot{M}$
(\ref{mjumu}), the problem of the orthogonalisation of vectors in $T_{f}$ (\ref{proor}) and so on.
All these problems have the data which depends smooth on the current state of $\Psi$, and,
consequently, on the time $t$. So, it is possible to solve these problems  with the help of the
perturbation theory and the methods of parametric continuation. These methods of the computational
linear algebra are widely used and their details are well-known, so we are not discussing it here
(\cite{Cont,Lin}.

\section{\textbf{Accuracy estimation and postprocessing in invariant manifolds constructing}}

Suppose that for the dynamical system (\ref{eqn1}) the approximate invariant manifold has been
constructed and the slow motion equations have been derived:

\begin{equation}\label{slag1}
\frac{dx_{sl}}{dt} = P_{x_{sl}}(J(x_{sl})), x_{sl}\in M,
\end{equation}
\noindent where $P_{x_{sl}}$ is the corresponding projector onto the tangent space $T_{x_{sl}}$ of
$M$. Suppose that we have solved the system (\ref{slag1}) and have obtained $x_{sl}(t)$. Let's
consider the following two questions:

\begin{itemize}
\item{How well this solution approximates the real solution $x(t)$ given the same initial conditions?}
\item{How is it possible to use the solution $x_{sl}(t)$ for it's refinement without solving the system (\ref{slag}) again?}
\end{itemize}

These two questions are interconnected. The first question states the problem of the {\it accuracy
estimation}. The second one states the problem of {\it postprocessing}.

The simplest (``naive") estimation is given by the ``invariance defect":

\begin{equation}\label{defag}
\Delta_{x_{sl}} = (1-P_{x_{sl}})J(x_{sl})
\end{equation}

\noindent compared with $J(x_{sl})$. For example, this estimation is given by $\epsilon =
\|\Delta_{x_{sl}}\|/\|J(x_{sl})\|$ using some appropriate norm.

Probably, the most comprehensive answer to this question can be given by solving the following
equation:

\begin{equation}\label{varia}
\frac{d(\delta x)}{dt}=\Delta_{x_{sl}(t)}+D_xJ(x)|_{x_{sl}(t)}\delta x.
\end{equation}

This linear equation describes the dynamics of the deviation $\delta x(t) = x(t) - x_{sl}(t)$ using
the linear approximation. The solution with zero initial conditions $\delta x(0) = 0$ allows
estimating $x_{sl}$ robustness as well as the error value. Substituting $x_{sl}(t)$ for
$x_{sl}(t)+\delta x(t)$ gives the required solution refinement. This {\it dynamical postprocessing}
\cite{MaTiWyPP} allows to refine the solution substantially and to estimate it's accuracy and
robustness. However, the price for this is solving the equation (\ref{varia}) with variable
coefficients. Thus, this dynamical postprocessing can be followed by a whole hierarchy of
simplifications, both dynamical and static. Let's mention some of them, starting from the dynamical
ones.

1) {\bf Freezing the coefficients}. In the equation (\ref{varia}) the linear operator
$D_xJ(x)|_{x_{sl}(t)}$ is replaced by it's value in some distinguished point $x^*$ (for example, in
the equilibrium) or it is frozen somehow else. As a result, one gets the equation with constant
coefficients and the explicit integration formula:

\begin{equation}\label{duam}
\delta x(t) = \int_0^t{exp(D^*(t-\tau))\Delta_{x_{sl}(\tau)}d\tau},
\end{equation}

\noindent where $D^*$ is the ``frozen" operator and $\delta x(0)=0$.

Another important way of freezing is substituting (\ref{varia}) for some {\it model equation}, i.e.
substituting $D_xJ(x)$ for $-\frac{1}{\tau^*}$, where $\tau^*$ is the relaxation time. In this case
the formula for $\delta x(t)$ has a very simple form:

\begin{equation}\label{duam1}
\delta x(t) = \int_0^t{e^{\frac{\tau-t}{\tau^*}}\Delta_{x_{sl}(\tau)}d\tau}.
\end{equation}

2) {\bf One-dimensional Galerkin-type approximation.} Another ``scalar" approximation is given by
projecting (\ref{varia}) on $\Delta(t)= \Delta_{x_{sl}(t)}$:

\begin{equation}\label{1var}
\delta x(t) = \delta(t)\cdot \Delta(t), \; \frac{d\delta(t)}{dt} =
1+\delta\frac{\langle\Delta|D\Delta\rangle -
\langle\Delta|\dot{\Delta}\rangle}{\langle\Delta|\Delta\rangle},
\end{equation}

\noindent where $\langle\hspace{1pt}|\rangle$ is an appropriate scalar product which can depend on
the point $x_{sl}$ (for example, the entropic scalar product), $D=D_xJ(x)|_{x_{sl}(t)}$ or the
self-adjoint linearization of this operator, or some approximation of it.

The ``hybrid" between equations (\ref{1var}) and (\ref{varia}) has the simplest form (but is more
difficult for computation than eq. (\ref{1var})):

\begin{equation}\label{hybrid}
\frac{d(\delta
x)}{dt}=\Delta(t)+\frac{\langle\Delta|D\Delta\rangle}{\langle\Delta|\Delta\rangle}\delta x.
\end{equation}
\noindent Here one uses the normalized matrix element
$\frac{\langle\Delta|D\Delta\rangle}{\langle\Delta|\Delta\rangle}$ instead of the linear operator
$D=D_xJ(x)|_{x_{sl}(t)}$.

Both equations (\ref{1var}) and (\ref{hybrid}) can be solved explicitly:

\begin{eqnarray}
\delta(t)&=&\int_0^t d \tau \exp\left(\int_{\tau}^t k(\theta)d\theta \right), \\ \delta x(t)&=&
\int_0^t \Delta(\tau)d \tau \exp\left(\int_{\tau}^t k_1(\theta)d\theta \right),
\end{eqnarray}
\noindent where $k(t)=\frac{\langle\Delta|D\Delta\rangle -
\langle\Delta|\dot{\Delta}\rangle}{\langle\Delta|\Delta\rangle},$
$k_1(t)=\frac{\langle\Delta|D\Delta\rangle}{\langle\Delta|\Delta\rangle}.$

The projection of $\Delta_{x_{sl}}(t)$ on the slow motion is zero, hence, for post-processing
analysis of the slow motion, the one-dimensional model (\ref{1var}) should be supplemented by one
more iteration:
\begin{eqnarray}
{d(\delta x_{sl}(t)) \over dt} = \delta(t) P_{x_{sl}(t)}(D_xJ(x_{sl}(t)))(\Delta(t));  \nonumber \\
\delta x_{sl}(t)= \int_0^t \delta(\tau) P_{x_{sl}(\tau)}(D_xJ(x_{sl}(\tau)))(\Delta(\tau))d\tau.
\end{eqnarray}
where $\delta(t)$ is the solution of (\ref{1var}).

3) For a {\bf static postprocessing} one uses stationary points of dynamical equations (\ref{varia})
or their simplified versions (\ref{duam}),(\ref{1var}). Instead of (\ref{varia}) one gets:

\begin{equation}\label{stvar}
D_xJ(x)|_{x_{sl}(t)}\delta x = -\Delta_{x_{sl}(t)}
\end{equation}

\noindent with one additional condition, $P_{x_{sl}}\delta x=0$. This is exactly the iteration
equation of the Newton's method in solving the invariance equation.

The corresponding stationary problems for the model equations and for the projections of
(\ref{varia}) on $\Delta$ are evident. We only mention that in the projection on $\Delta$ one gets a
step of the relaxation method for the invariant manifold construction.

For the static postprocessing with frozen parameters the ``naive" estimation given by the
``invariance defect" (\ref{defag})  makes sense.

\section{\bf Example: Dumbell model, explosion of the Gaussian anzatz and polymer stretching in flow}

Here is an example of application of the thermodynamic projector method. In this example we consider
the following simplest one-dimensional kinetic equation for the configuration distribution function
$\Psi(q,t)$, where $q$ is the reduced vector connecting the beads of the dumbell. This equation is
slightly different from the FPE considered above. It is nonlinear, because of the dependence of $U$
on the moment $M_{2}[\Psi]=\int q^{2}\Psi(q) dq$. This dependence allows us to get the exact
quasiequilibrium equations on $M_{2}$, but this equations are not solving the problem: this
quasiequilibrium manifold may become unstable when the flow is present \cite{IK00}. Here is this
model:
\begin{equation}\label{530}
\partial_{t}\Psi=-\partial_{q}\{\alpha(t)q\Psi\}+\frac{1}{2}\partial^{2}_{q}\Psi.
\end{equation}
Here
\begin{equation}\label{531}
\alpha(t)=\kappa(t)-\frac{1}{2}f(M_{2}(t)),
\end{equation}
$\kappa(t)$ is the given time-independent velocity gradient, $t$ is the reduced time, and the
function $-fq$ is the reduced spring force. Function $f$ may depend on the second moment of the
distribution function $M_{2}=\int q^{2}\Psi(q,t)dq$. In particular, the case $f\equiv1$ corresponds
to the linear Hookean spring, while $f=[1-M_{2}(t)/b]^{-1}$ corresponds to the self-consistent finite
extension nonlinear elastic spring (the FENE-P model, first introduced in \cite{FENEP}). The second
moment $M_{2}$ occurs in the FENE-P force $f$ as the result of the pre-averaging approximation to the
original FENE model (with nonlinear spring force $f=[1-q^{2}/b]^{-1}$). Leading to closed
constitutive equations, the FENE-P model is frequently used in simulations of complex rheological
flows as the reference for more sophisticated closures to the FENE model \cite{HCO,Keu98,Martin}. The
parameter $b$ changes the characteristics of the force law from Hookean at small extensions to a
confining force for $q^{2}\rightarrow b$. Parameter $b$ is roughly equal to the number of monomer
units represented by the dumbell and should therefore be a large number. In the limit
$b\rightarrow\infty$, the Hookean spring is recovered. Recently, it has been demonstrated that FENE-P
model appears as first approximation within a systematic self-confident expansion of nonlinear forces
\cite{IKOe99,GKIOeNONNEWT2001}.

Equation (\ref{530}) describes an ensemble of non-interacting dumbells subject to a
pseudo-elongational flow with fixed kinematics. As is well known, the Gaussian distribution function,
\begin{equation}\label{532}
\Psi^{G}(M_{2})=\frac{1}{\sqrt{2\pi M_{2}}}\exp\left[-\frac{q^{2}}{2M_{2}}\right],
\end{equation}
solves equation (\ref{530}) provided the second moment $M_{2}$ satisfies
\begin{equation}\label{533}
\frac{dM_{2}}{dt}=1+2\alpha(t)M_{2}.
\end{equation}
Solution (\ref{532}) and (\ref{533}) is the valid macroscopic description if all other solutions of
the equation (\ref{530}) are rapidly attracted to the family of Gaussian distributions (\ref{532}).
In other words \cite{GKTTSP94}, the special solution (\ref{532}) and (\ref{533}) is the macroscopic
description if equation (\ref{532}) is the stable invariant manifold of the kinetic equation
(\ref{530}). If not, then the Gaussian solution is just a member of the family of solutions, and
equation (\ref{533}) has no meaning of the macroscopic equation. Thus, the complete answer to the
question of validity of the equation (\ref{533}) as the macroscopic equation requires a study of
dynamics in the neighborhood of the manifold (\ref{532}). Because of the simplicity of the model
(\ref{530}), this is possible to a satisfactory level even for $M_{2}$-dependent spring forces.

In the paper \cite{IK00} it was shown, that there is a possibility of ``explosion" of the Gaussian
manifold: with the small initial deviation from it, the solutions of the equation (\ref{530}) are
very fast going far from, and then slowly come back to the stationary point which is located on the
Gaussian manifold. The distribution function $\Psi$ is stretched fast, but looses the Gaussian form,
and after that the Gaussian form recovers slowly with the new value of $M_{2}$. Let us describe
briefly the results of \cite{IK00}.

Let $M_{2n}=\int q^{2n}\Psi dq$ denote the even moments (odd moments vanish by symmetry). We consider
deviations $\mu_{2n}=M_{2n}-M_{2n}^{\rm G}$, where $M_{2n}^{\rm G}=\int q^{2n} \Psi^{\rm G}dq$ are
moments of the Gaussian distribution function (\ref{532}). Let $\Psi(q,t_0)$ be the initial condition
to the Eq.\ (\ref{530}) at time $t=t_0$. Introducing functions,
\begin{equation}
\label{result0} p_{2n}(t,t_0)=\exp\left[4n\int_{t_0}^{t}\alpha(t')dt'\right],
\end{equation}
where $t\ge t_0$, and $2n \ge 2$, the {\it exact} time evolution of the deviations $\mu_{2n}$ for
$2n\ge 2$ reads
\begin{equation} \label{result1}
    \mu_4(t)=p_4(t,t_0)\mu_4(t_0),
\end{equation}
and
\begin{equation} \label{result2}
    \mu_{2n}(t)=\left[ \mu_{2n}(t_0) + 2n(4n-1)\int_{t_0}^t
    \mu_{2n-2}(t')p_{2n}^{-1}(t',t_0)dt' \right] p_{2n}(t,t_0),
\end{equation}
for $2n\ge 3$. Equations (\ref{result0}), (\ref{result1}) and (\ref{result2}) describe evolution near
the Gaussian solution for arbitrary initial condition $\Psi(q,t_0)$. Notice that explicit evaluation
of the integral in the Eq.\ (\ref{result0}) requires solution to the moment equation (\ref{533})
which is not available in the analytical form for the FENE-P model.

It is straightforward to conclude that any solution with a non-Gaussian initial condition converges
to the Gaussian solution asymptotically as $t\to\infty$ if

\begin{equation}
\label{result3} \lim_{t\to\infty}\int_{t_0}^t\alpha(t')dt'<0.
\end{equation}
However, even if this asymptotic condition is met, deviations from the Gaussian solution may survive
for considerable {\it finite} times. For example, if for some finite time $T$, the integral in the
Eq.\ (\ref{result0}) is estimated as $\int_{t_0}^t\alpha(t')dt'>\alpha (t-t_0)$, $\alpha>0$, $t\le
T$, then the Gaussian solution becomes exponentially unstable during this time interval. If this is
the case, the moment equation (\ref{533}) cannot be regarded as the macroscopic equation. Let us
consider specific examples.

For the Hookean spring ($f\equiv 1$) under a constant elongation ($\kappa={\rm const}$), the Gaussian
solution is exponentially stable for $\kappa<0.5$, and it becomes exponentially unstable for
$\kappa>0.5$. The exponential instability in this case is accompanied by the well known breakdown of
the solution to the Eq.\ (\ref{533}) due to infinite stretching of the dumbbell. Similar instability
has been found numerically in three-dimensional flows for high Weissenberg numbers
\cite{PGA851,PGA852}.

\begin{figure}[t]
\centering{
\includegraphics[width=130mm,height=120mm]{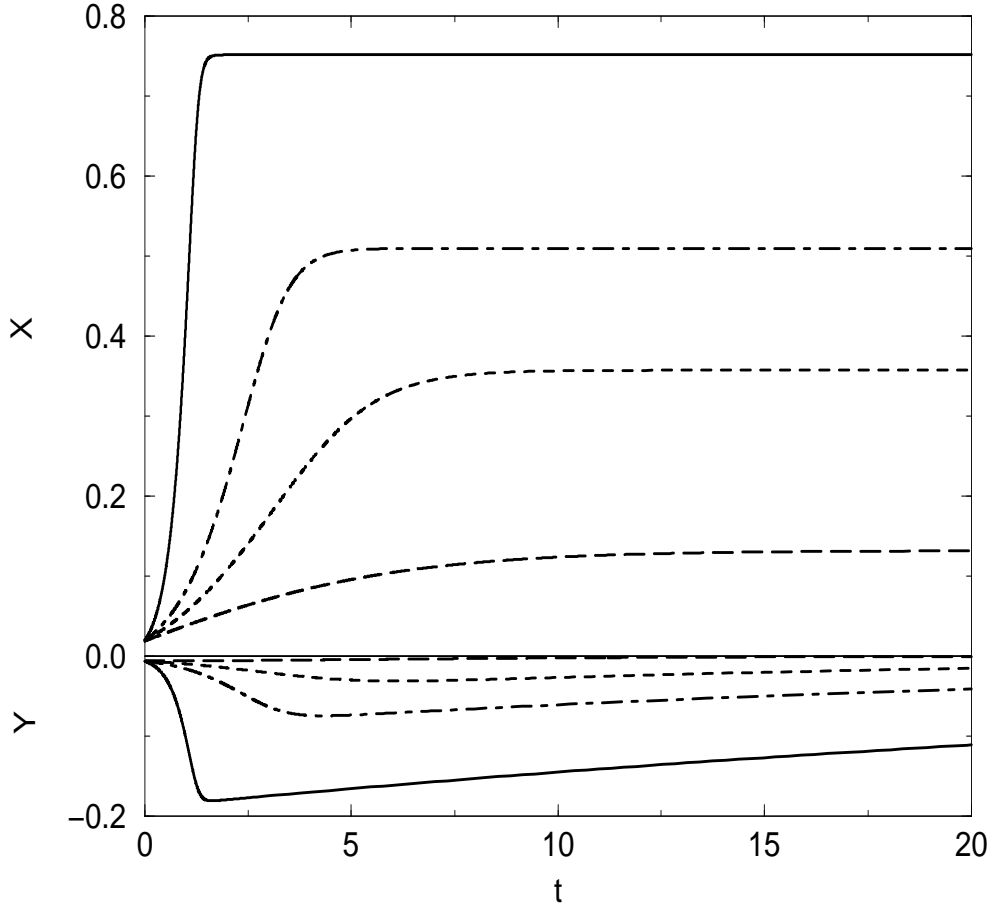}
\caption{Deviations of reduced moments from the Gaussian solution as a function of reduced time $t$
in pseudo-elongation flow for the FENE-P model. Upper part: Reduced second moment $X=M_2/b$. Lower
part: Reduced deviation of fourth moment from Gaussian solution $Y=-\mu_4^{1/2}/b$. Solid:
$\kappa=2$, dash-dot: $\kappa=1$, dash: $\kappa=0.75$, long dash: $\kappa=0.5$. (The figure from the
paper \cite{IK00}, computed by P. Ilg.)} \label{EPJ713_fig}}
\end{figure}

Eqs.\ (\ref{533}) and (\ref{result1}) were integrated  by the 5-th order Runge-Kutta method with
adaptive time step. The FENE-P parameter $b$ was set equal to 50. The initial condition was
$\Psi(q,0)=C(1-q^2/b)^{b/2}$, where $C$ is the normalization (the equilibrium of the FENE model,
notoriously close to the FENE-P equilibrium \cite{Herrchen}). For this initial condition, in
particular, $\mu_4(0)=-6b^2/[(b+3)^2(b+5)]$ which is about 4$\%$ of the value of $M_4$ in the
Gaussian equilibrium for $b=50$. In Fig.~\ref{EPJ713_fig} we demonstrate deviation $\mu_4(t)$ as a
function of time for several values of the flow. Function $M_2(t)$ is also given for comparison. For
small enough $\kappa$ we find an adiabatic regime, that is $\mu_4$ relaxes exponentially to zero. For
stronger flows, we observe an initial {\it fast runaway} from the invariant manifold with $|\mu_4|$
growing over three orders of magnitude compared to its initial value. After the maximum deviation has
been reached, $\mu_4$ relaxes to zero. This relaxation is exponential as soon as the solution to Eq.\
(\ref{533}) approaches the steady state. However, the time constant for this exponential relaxation
$|\alpha_{\infty}|$ is very small. Specifically, for large $\kappa$,
\begin{equation}
\label{alpha_lim} \alpha_{\infty}=\lim_{t\to\infty}\alpha(t)=-\frac{1}{2b}+O(\kappa^{-1}).
\end{equation}
Thus, the steady state solution is unique and Gaussian but the stronger is the flow, the larger is
the initial runaway from the Gaussian solution, while the return to it thereafter becomes
flow-independent. Our observation demonstrates that, though the stability condition (\ref{result3})
is met, {\it significant deviations from the Gaussian solution persist over the times when the
solution of Eq.}\ (\ref{533}) {\it is already reasonably close to the stationary state.} If we accept
the usually quoted physically reasonable minimal value of parameter $b$ of the order $20$ then the
minimal relaxation time is of order $40$ in the reduced time units of Fig.~\ref{EPJ713_fig}. We
should also stress that the two limits, $\kappa\to\infty$ and $b\to\infty$, are not commutative, thus
it is not surprising that the estimation (\ref{alpha_lim}) does not reduce to the above mentioned
Hookean result as $b\to\infty$. Finally, peculiarities of convergence to the Gaussian solution are
even furthered if we consider more complicated (in particular, oscillating) flows $\kappa(t)$.

In accordance with \cite{IK00} the anzatz for $\Psi$ can be suggested in the following form:
\begin{equation}\label{Anz}
\Psi^{An}(\{\sigma,\varsigma\},q)=\frac{1}{2\sigma\sqrt{2\pi}}\left(e^{-\frac{(q+\varsigma)^{2}}{2\sigma^{2}}}+e^{-\frac{(q-\varsigma)^{2}}{2\sigma^{2}}}\right).
\end{equation}
Natural inner coordinates on this manifold are $\sigma$ and $\varsigma$. Note, that now
$\sigma^{2}\neq M_{2}$. The value $\sigma^{2}$ is a dispersion of one of the Gaussian summands in
(\ref{Anz}),
\begin{eqnarray*}
M_{2}(\Psi^{An}(\{\sigma,\varsigma\},q))=\sigma^{2}+\varsigma^{2}.
\end{eqnarray*}
To build the thermodynamic projector on the manifold (\ref{Anz}), the thermodynamic Lyapunov function
is necessary. It is necessary to emphasize, that equations (\ref{530}) are nonlinear. For such
equations, the arbitrarity in the choice of the thermodynamic Lyapunov function is much smaller.
Nevertheless, such function exists. It is the free energy
\begin{equation}\label{Free}
F=U(M_{2}[\Psi])-TS[\Psi],
\end{equation}
where
\begin{eqnarray*}
S[\Psi]=-\int\Psi(\ln\Psi-1)dq,
\end{eqnarray*}
$U(M_{2}[\Psi])$ is the potential energy in the mean field approximation, $T$ is the temperature
(further we assume that $T=1$). The thermodynamic properties of the mean field models in polymer
physics are studied in the recent paper \cite{MaKaHa2003}

Note, that Kullback-form entropy $S_{k}=-\int\Psi\ln\left(\frac{\Psi}{\Psi^{*}}\right)$ also has the
form $S_{k}=-F/T$:
\begin{eqnarray*}
\Psi^{*}=\exp(-U),\\ S_{k}[\Psi]=-\langle U\rangle-\int\Psi\ln\Psi dq.
\end{eqnarray*}
If $U(M_{2}[\Psi])$ in the mean field approximation is the convex function of $M_{2}$, then the free
energy (\ref{Free}) is the convex functional too.

For the FENE-P model $U=-\ln[1-M_{2}/b]$.

In accordance to the thermodynamics the vector of flow of $\Psi$ must be proportional to the gradient
of the corresponding chemical potential $\mu$:
\begin{equation}\label{Flux}
J=-B(\Psi)\nabla_{q}\mu,
\end{equation}
where $\mu=\frac{\delta F}{\delta\Psi}$, $B\geq0$. From the equation (\ref{Free}) it follows, that
\begin{eqnarray}\label{muflux}
\mu=\frac{d U(M_{2})}{d M_{2}}\cdot q^{2}+\ln\Psi\nonumber\\ J=-B(\Psi)\left[2\frac{dU}{dM_{2}}\cdot
q+\Psi^{-1}\nabla_{q}\Psi\right].
\end{eqnarray}
If we suppose here $B=\frac{D}{2}\Psi$, then we get
\begin{eqnarray}\label{TDeq}
J=-D\left[\frac{dU}{dM_{2}}\cdot q\Psi+\frac{1}{2}\nabla_{q}\Psi\right]\nonumber\\
\frac{\partial\Psi}{\partial t}=div_{q}J=D\frac{d U(M_{2})}{d
M_{2}}\partial_{q}(q\Psi)+\frac{D}{2}\partial^{2}q\Psi,
\end{eqnarray}
When $D=1$ this equations coincide with (\ref{530}) in the absence of the flow: due to equation
(\ref{TDeq}) $dF/dt\leq0$.

Let us construct the thermodynamic projector with the help of the thermodynamic Lyapunov function $F$
(\ref{Free}). Corresponding entropic scalar product in the point $\Psi$ has the form
\begin{equation}\label{Scal}
\left.\langle f|g\rangle=\frac{d^{2}U}{dM_{2}^{2}}\right|_{M_{2}=M_{2}[\Psi]}\cdot\int
q^{2}f(q)dq\cdot\int q^{2}g(q)dq+\int\frac{f(q)g(q)}{\Psi(q)}dq
\end{equation}
During the investigation of the anzatz (\ref{Anz}) the scalar product (\ref{Scal}), constructed for
the corresponding point of the Gaussian manifold with $M_{2}=\sigma^{2}$, will be used. It will let
us to investigate the neighborhood of the Gaussian manifold (and to get all the results in the
analytical form):
\begin{equation}\label{ScalG}
\left.\langle f|g\rangle_{\sigma^{2}}=\frac{d^{2}U}{dM_{2}^{2}}\right|_{M_{2}=\sigma^{2}}\cdot\int
q^{2}f(q)dq\cdot\int q^{2}g(q)dq+\sigma\sqrt{2\pi}\int e^{\frac{q^{2}}{2\sigma^{2}}}f(q)g(q)dq
\end{equation}
Also we will need to know the functional $DF$ in the point of Gaussian manifold:
\begin{equation}\label{Prod}
\left.DF_{\sigma^{2}}(f)=\left(\frac{d U(M_{2})}{dM_{2}}\right|_{M_{2}=\sigma^{2}}
-\frac{1}{2\sigma^{2}}\right)\int q^{2}f(q)dq,
\end{equation}
\noindent (with the condition $\int f(q)dq=0$). The point
\begin{eqnarray*}
\left.\frac{d U(M_{2})}{dM_{2}}\right|_{M_{2}=\sigma^{2}}=\frac{1}{2\sigma^{2}},
\end{eqnarray*}
corresponds to the equilibrium.

The tangent space to the manifold (\ref{Anz}) is spanned by the vectors
\begin{eqnarray}\label{basis}
&&f_{\sigma}=\frac{\partial\Psi^{An}}{\partial(\sigma^{2})}; \:
f_{\varsigma}=\frac{\partial\Psi^{An}}{\partial(\varsigma^{2})};\nonumber\\
f_{\sigma}&=&\frac{1}{4\sigma^{3}\sqrt{2\pi}}\left[e^{-\frac{(q+\varsigma)^{2}}{2\sigma^{2}}}
\frac{(q+\varsigma)^{2}-\sigma^{2}}{\sigma^{2}}+e^{-\frac{(q-\varsigma)^{2}}{2\sigma^{2}}}
\frac{(q-\varsigma)^{2}-\sigma^{2}}{\sigma^{2}} \right];\\
f_{\varsigma}&=&\frac{1}{4\sigma^{2}\varsigma\sqrt{2\pi}}\left[-e^{-\frac{(q+\varsigma)^{2}}{2\sigma^{2}}}
\frac{q+\varsigma}{\sigma}+e^{-\frac{(q-\varsigma)^{2}}{2\sigma^{2}}} \frac{(q-\varsigma)}{\sigma}
\right];\nonumber
\end{eqnarray}
The Gaussian entropy (free energy) production in the directions $f_{\sigma}$ and $f_{\varsigma}$
(\ref{Prod}) has a very simple form:
\begin{eqnarray}\label{Fpro}
\left.DF_{\sigma^{2}}(f_{\varsigma})=DF_{\sigma^{2}}(f_{\sigma})=\frac{d
U(M_{2})}{dM_{2}}\right|_{M_{2}=\sigma^{2}}-\frac{1}{2\sigma^{2}}.
\end{eqnarray}
The linear subspace $\ker DF_{\sigma^{2}}$ in $lin\{f_{\sigma},f_{\varsigma}\}$ is spanned by the
vector $f_{\varsigma}-f_{\sigma}$.

Let us have the given vector field $d\Psi/dt=\Phi(\Psi)$ in the point $\Psi(\{\sigma,\varsigma\})$.
We need to build the projection of $\Phi$ onto the tangent space $T_{\sigma,\varsigma}$ in the point
$\Psi(\{\sigma,\varsigma\})$:
\begin{equation}\label{Prosigma}
P^{th}_{\sigma,\varsigma}(\Phi)=\varphi_{\sigma}f_{\sigma}+\varphi_{\varsigma}f_{\varsigma}.
\end{equation}
This equation means, that the equations for $\sigma^{2}$ and $\varsigma^{2}$ will have the form
\begin{equation}\label{eqsigma}
\frac{d\sigma^{2}}{dt}=\varphi_{\sigma};\:\: \frac{d\varsigma^{2}}{dt}=\varphi_{\varsigma}
\end{equation}
Projection $(\varphi_{\sigma},\varphi_{\varsigma})$ can be found from the following two equations:
\begin{eqnarray}\label{psieq}
\varphi_{\sigma}+\varphi_{\varsigma}=\int q^{2}\Phi(\Psi)(q)dq\nonumber;\\
\langle\varphi_{\sigma}f_{\sigma}+\varphi_{\varsigma}f_{\varsigma}|f_{\sigma}-f_{\varsigma}\rangle_{\sigma^{2}}
=\langle\Phi(\Psi)|f_{\sigma}-f_{\varsigma}\rangle_{\sigma^{2}},
\end{eqnarray}
where $\langle
f|g\rangle_{\sigma^{2}}=\langle\Phi(\Psi)|f_{\sigma}-f_{\varsigma}\rangle_{\sigma^{2}}$,
(\ref{Scal}). First equation of (\ref{psieq}) means, that the time derivative $dM_{2}/dt$ is the same
for the initial and the reduced equations. Due to the formula for the dissipation of the free energy
(\ref{Prod}), this equality is equivalent to the persistence of the dissipation in the neighborhood
of the Gaussian manifold.

The second equation in (\ref{psieq}) means, that $\Phi$ is projected orthogonally on $\ker DS\bigcap
T_{\sigma,\varsigma}$. Let us use the orthogonality with respect to the entropic scalar product
(\ref{ScalG}). The solution of equations (\ref{psieq}) has the form
\begin{eqnarray}\label{projphi}
\frac{d\sigma^{2}}{dt}=\varphi_{\sigma}=\frac{\langle\Phi|f_{\sigma}-f_{\varsigma}\rangle_{\sigma^{2}}+M_{2}(\Phi)(\langle
f_{\varsigma}|f_{\varsigma}\rangle_{\sigma^{2}}-\langle
f_{\sigma}|f_{\varsigma}\rangle_{\sigma^{2}})}{\langle
f_{\sigma}-f_{\varsigma}|f_{\sigma}-f_{\varsigma}\rangle_{\sigma^{2}}}\nonumber,\\\\
\frac{d\varsigma^{2}}{dt}=\varphi_{\varsigma}=\frac{-\langle\Phi|f_{\sigma}-f_{\varsigma}\rangle_{\sigma^{2}}+M_{2}(\Phi)(\langle
f_{\sigma}|f_{\sigma}\rangle_{\sigma^{2}}-\langle
f_{\sigma}|f_{\varsigma}\rangle_{\sigma^{2}})}{\langle
f_{\sigma}-f_{\varsigma}|f_{\sigma}-f_{\varsigma}\rangle_{\sigma^{2}}}\nonumber,
\end{eqnarray}
where $\Phi=\Phi(\Psi)$, $M_{2}(\Phi)=\int q^{2}\Phi(\Psi)dq$.

It is easy to check, that the formulas (\ref{projphi}) are indeed defining the projector: if
$f_{\sigma}$ (or $f_{\varsigma}$) is substituted there instead of the function $\Phi$, then we will
get $\varphi_{\sigma}=1, \varphi_{\varsigma}=0$ (or $\varphi_{\sigma}=0, \varphi_{\varsigma}=1$,
respectively). Let us substitute the right part of the initial kinetic equations (\ref{530}),
calculated in the point $\Psi(q)=\Psi(\{\sigma,\varsigma\},q)$ (see the equation (\ref{Anz})) in the
equation (\ref{projphi}) instead of $\Phi$. We will get the closed system of equations on
$\sigma^{2}, \varsigma^{2}$ in the neighborhood of the Gaussian manifold.

This system describes the dynamics of the distribution function $\Psi$. The distribution function is
represented as the half-sum of two Gaussian distributions with the averages of distribution
$\pm\varsigma$ and mean-square deviations $\sigma$. All integrals in the right-hand part of
(\ref{projphi}) are possible to calculate analytically.

Basis $(f_{\sigma},f_{\varsigma})$ is convenient to use everywhere, except the points in the Gaussian
manifold, $\varsigma=0$, because if $\varsigma\rightarrow0$, then
\begin{eqnarray*}
f_{\sigma}-f_{\varsigma}=O\left(\frac{\sigma^{2}}{\varsigma^{2}}\right)\rightarrow0.
\end{eqnarray*}
To analyze the relaxation in the small neighborhood of the Gaussian manifold it is more convenient to
use another basis:
\begin{eqnarray*}
F^{+}=f_{\sigma}+f_{\varsigma}\\ F^{-}=\frac{\sigma^{2}}{\varsigma^{2}}(f_{\sigma}-f_{\varsigma}).
\end{eqnarray*}
It corresponds to a reparametrization of the initial manifold (\ref{Anz}):
\begin{equation}\label{Anz1}
\Psi(\{\xi,\varsigma\},q)=\frac{1}{2\sqrt{2\pi}\sqrt{\xi^{2}-\varsigma^{2}}}\left(e^{-\frac{(q+\varsigma)^{2}}{2(\xi^{2}-\varsigma^{2})}}+e^{-\frac{(q-\varsigma)^{2}}{2(\xi^{2}-\varsigma^{2})}}\right).
\end{equation}
Let us analyze the stability of the Gaussian manifold to the ``dissociation" of the Gaussian peak in
two peaks (\ref{Anz}). To do this, it is necessary to find first nonzero term in the Taylor expansion
in $\varsigma^{2}$ of the right-hand side of the second equation in the system (\ref{projphi}). The
denominator has the order of $\varsigma^{4}$, the numerator has, as it is easy to see, the order not
less, than $\varsigma^{6}$ (because the Gaussian manifold is invariant with respect to the initial
system).

Let us denote $G_{\sigma}=\frac{1}{\sqrt{2\pi}}e^{-\frac{q^{2}}{\sigma^{2}}}$. Then we get
\begin{eqnarray*}
\Psi(\{\sigma,\varsigma\},q)=G_{\sigma}(q)\left[1+\frac{1}{2}\frac{\varsigma^{2}}{\sigma^{2}}\left(\frac{q^{2}}{\sigma^{2}}-1\right)+\frac{1}{4}\frac{\varsigma^{4}}{\sigma^{4}}\left(\frac{1}{2}-\frac{q^{2}}{\sigma^{2}}+\frac{1}{6}\frac{q^{4}}{\sigma^{4}}\right)\right]+o\left(\frac{\varsigma^{4}}{\sigma^{4}}\right);\\
f_{\sigma}=\frac{G_{\sigma}(q)}{2\sigma^{2}}\left[\frac{q^{2}}{\sigma^{2}}-1+\frac{\varsigma^{2}}{\sigma^{2}}\left(\frac{1}{2}\frac{q^{4}}{\sigma^{4}}-3\frac{q^{2}}{\sigma^{2}}+\frac{3}{2}\right)+\frac{\varsigma^{4}}{\sigma^{4}}\left(\frac{1}{24}\frac{q^{6}}{\sigma^{6}}-\frac{15}{24}\frac{q^{4}}{\sigma^{4}}+\frac{15}{8}\frac{q^{2}}{\sigma^{2}}-\frac{5}{8}\right)\right]+o\left(\frac{\varsigma^{4}}{\sigma^{4}}\right);\\
f_{\varsigma}=\frac{G_{\sigma}(q)}{2\sigma^{2}}\left[\frac{q^{2}}{\sigma^{2}}-1+\frac{\varsigma^{2}}{\sigma^{2}}\left(\frac{1}{6}\frac{q^{4}}{\sigma^{4}}-\frac{q^{2}}{\sigma^{2}}+\frac{1}{2}\right)+\frac{\varsigma^{4}}{\sigma^{4}}\left(\frac{1}{120}\frac{q^{6}}{\sigma^{6}}-\frac{1}{8}\frac{q^{4}}{\sigma^{4}}+\frac{3}{8}\frac{q^{2}}{\sigma^{2}}-\frac{1}{8}\right)\right]+o\left(\frac{\varsigma^{4}}{\sigma^{4}}\right);\\
f_{\sigma}-f_{\varsigma}=\frac{\varsigma^{2}}{\sigma^{2}}\frac{1}{2\sigma^{2}}G_{\sigma}(q)\left[\frac{1}{3}\frac{q^{4}}{\sigma^{4}}-2\frac{q^{2}}{\sigma^{2}}+1+\frac{\varsigma^{2}}{\sigma^{2}}\left(\frac{1}{30}\frac{q^{6}}{\sigma^{6}}-\frac{1}{2}\frac{q^{4}}{\sigma^{4}}+\frac{3}{2}\frac{q^{2}}{\sigma^{2}}-\frac{1}{2}\right)\right]+o\left(\frac{\varsigma^{4}}{\sigma^{4}}\right).
\end{eqnarray*}
Let us calculate $\partial_{t}\Psi=\Phi(\Psi(\{\sigma,\varsigma\}))$ with the accuracy up to
$\varsigma^{4}$:
\begin{eqnarray*}
\frac{1}{2}\partial^{2}_{q}\Psi(\{\sigma,\varsigma\})=f_{\sigma};\\
M_{2}(\frac{1}{2}\partial^{2}_{q}\Psi(\{\sigma,\varsigma\}))=1;\\
M_{2}(\Psi(\{\sigma,\varsigma\}))=\sigma^{2}+\varsigma^{2};\\
-\alpha\partial_{q}(q\Psi(\{\sigma,\varsigma\}))=\alpha
G_{\sigma}(q)\left[\frac{q^{2}}{\sigma^{2}}-1+\frac{\varsigma^{2}}{\sigma^{2}}\left(\frac{1}{2}\frac{q^{4}}{\sigma^{4}}-2\frac{q^{2}}{\sigma^{2}}+\frac{1}{2}\right)+\right.\\
\left.\frac{\varsigma^{4}}{\sigma^{4}}\left(\frac{1}{24}\frac{q^{6}}{\sigma^{6}}-\frac{11}{24}\frac{q^{4}}{\sigma^{4}}+\frac{7}{8}\frac{q^{2}}{\sigma^{2}}-\frac{1}{8}\right)\right]+o\left(\frac{\varsigma^{4}}{\sigma^{4}}\right)\\
M_{2}(-\alpha\partial_{q}(q\Psi(\{\sigma,\varsigma\})))=2\alpha(\sigma^{2}+\varsigma^{2})+o\left(\frac{\varsigma^{4}}{\sigma^{4}}\right).
\end{eqnarray*}
The diffusion part gives the zero contribution to the numerator of the equation (\ref{projphi}):
\begin{eqnarray*}
-\langle f_{\sigma}|f_{\sigma}-f_{\varsigma}\rangle+\langle
f_{\sigma}|f_{\sigma}-f_{\varsigma}\rangle=0,
\end{eqnarray*}
therefore to find $d\varsigma/dt$ it is sufficient to use $\Phi_{1}=-\alpha\partial_{q}(q\Psi)$, so
we get
\begin{eqnarray*}
M_{2}(\Phi_{1}(\Psi(\{\sigma,\varsigma\})))f_{\sigma}-\Phi_{1}(\Psi(\{\sigma,\varsigma\}))=\alpha
G_{\sigma}(q)\frac{\varsigma^{4}}{\sigma^{4}}\left(\frac{1}{3}\frac{q^{4}}{\sigma^{4}}-2\frac{q^{2}}{\sigma^{2}}+1\right)+o\left(\frac{\varsigma^{4}}{\sigma^{4}}\right)\\
=2\alpha\sigma^{2}\frac{\varsigma^{2}}{\sigma^{2}}(f_{\sigma}-f_{\varsigma})+o\left(\frac{\varsigma^{4}}{\sigma^{4}}\right).
\end{eqnarray*}
Thus
\begin{equation}\label{itog}
\frac{1}{\sigma^{2}}\frac{d\varsigma^{2}}{dt}=2\alpha\frac{\varsigma^{2}}{\sigma^{2}}+o\left(\frac{\varsigma^{4}}{\sigma^{4}}\right).
\end{equation}
So, if $\alpha>0$, then $\varsigma^{2}$ grows exponentially ($\varsigma\sim e^{\alpha t}$) and the
Gaussian manifold is unstable; if $\alpha<0$, then $\varsigma^{2}$ decreases exponentially and the
Gaussian manifold is stable.

\begin{figure}[t]
\begin{centering}
\includegraphics[width=160mm, height=127mm]{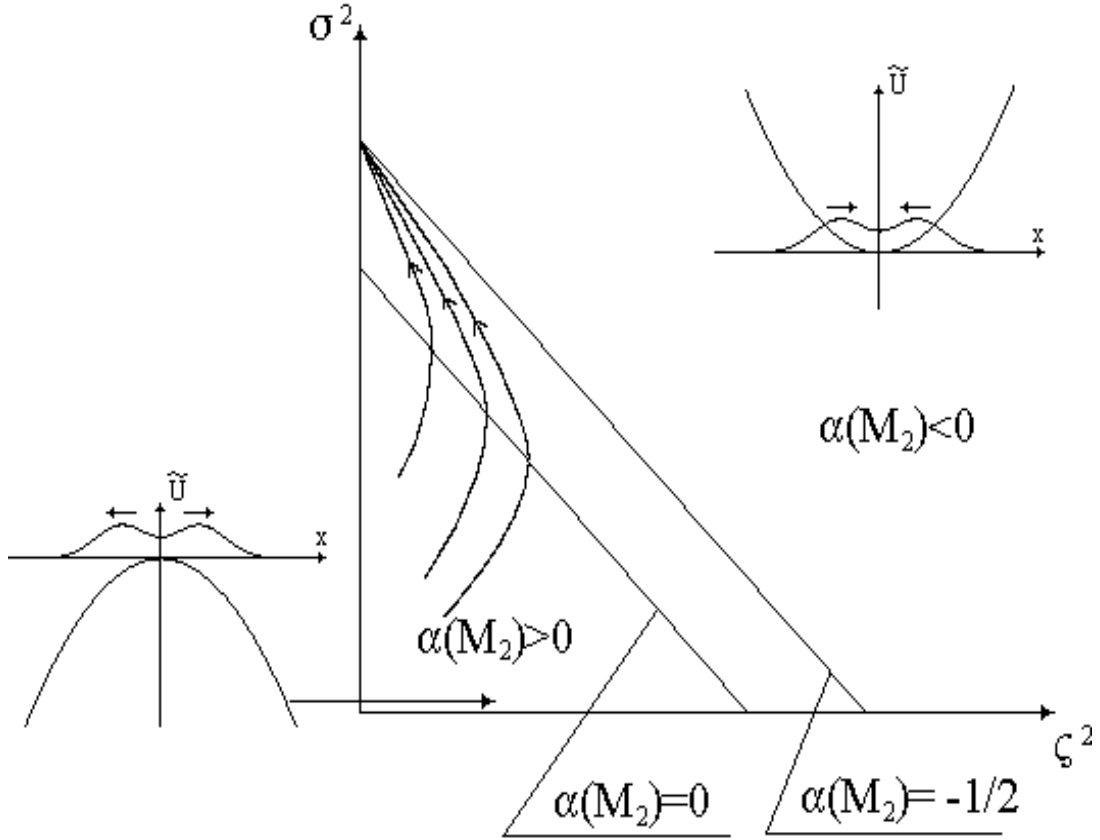}
\caption {Phase trajectories for two-peak approximation, FENE-P model. The vertical axis
($\varsigma=0$) corresponds to the Gaussian manifold. The triangle with $\alpha(M_2)>0$ is the domain
of exponential instability. }
    \label{figFENEP}
\end{centering}
\end{figure}

The form of the phase trajectories is shown qualitative on the figure \ref{figFENEP}.

\begin{figure}[t]
\begin{centering}
\includegraphics[width=160mm, height=87mm]{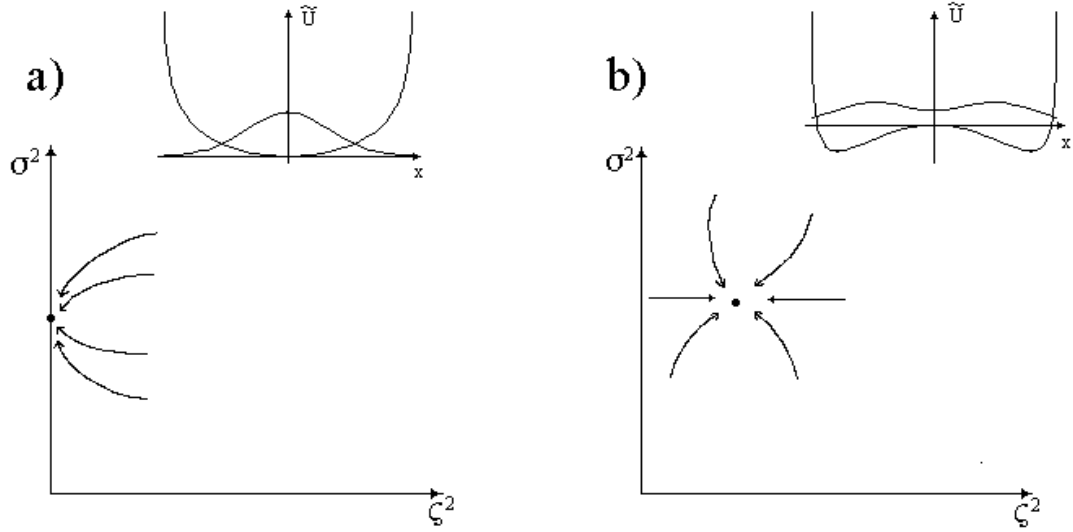}
\caption {Phase trajectories for two-peak approximation, FENE model: {\bf a)} A stable equilibrium on
the vertical axis, one stable peak; {\bf b)} A stable equilibrium with $\varsigma>0$, stable two-peak
configuration.}
    \label{figFENE}
\end{centering}
\end{figure}

For the real equation FPE (for example, with the FENE potential) the motion in presence of the flow
can be represented as the motion in the effective potential well $\tilde{U}(q)=U(q)-\kappa q^{2}$.
Different variants of the phase portrait for the FENE potential are present on the figure
\ref{figFENE}. Instability and dissociation of the unimodal distribution functions (``peaks") for the
FPE is the general effect when the flow is present.

The instability occurs when the matrix $\partial^{2}\tilde{U}/\partial q_{i}\partial q_{j}$ starts to
have negative eigenvalues ($\tilde{U}$ is the effective potential energy,
$\tilde{U}(q)=U(q)-\sum_{i,j}\kappa_{i,j}q_{i}q_{j}$).

The stationary polymodal distribution corresponds to the persistence of several local minima of the
function $\tilde{U}(q)$. The multidimensional case is different from one-dimensional because it has
the huge amount of possible configurations. All normal forms of the catastrophe of ``birth of the
critical point" are well investigated and known \cite{ArVarGZ1995-1998}. Every dissociation of the
peak is connected with such a catastrophe. The number of the new peaks is equal to the number of the
new local minima of $U$.

It is not very difficult to perform the analysis of the equations (\ref{projphi}) for every quantity
of peaks and every potential. Moreover, for the polynomial potentials all the necessary integrals are
possible to calculate analytically (if the coefficients of the scalar product and entropy production
are taken in the Gaussian point). The same situation is also for the general Gaussian distributions:
\begin{eqnarray*}
G_{\Sigma}=const\cdot\exp\left(-\frac{1}{2}\sum_{i,j}(\Sigma^{-1})_{ij}q_{i}q_{j}\right),
\end{eqnarray*}
where $\Sigma$ is the covariance matrix. Here in the equation for the effective energy we have the
symmetric part of the tensor $\kappa_{ij}={\partial^2 U}/{\partial q_{i}\partial q_{j}}$. The
presence of the unsimmetric part may lead to the relaxation oscillations (both for FPE and for the
peak dynamics).

For the modeling of dynamics of the multimodal distributions for FPE with the presence of the flow
(the flow may be nonstationary) it seems to be useful to use the physically clear modeling of the
distribution function as a sum of the finite number of the Gaussian peaks. Thermodynamic projector
gives us an opportunity to make this models thermodynamically consistent.

\section{\bf Conclusion}

In this work we presented a toolbox for the development and reduction of the dynamical models of
nonequilibrium systems with the persistence of the correct dissipation.

The basic notions of this toolbox are: entropy, quasiequilibrium (MaxEnt) distribution, dual
variables, thermodynamic projector.

The main technical ideas are: Legendre Integrators, dynamical postprocessing, transformation of
almost arbitrary anzatz to a thermodynamically consistent model via thermodynamic projector.

The Legendre Integrators are based on a simple, but very useful idea: to write and solve dynamic
equations for dual variables. This idea is efficient, because to obtain the dynamic equations for
dual variables it is necessary to solve linear equations. To get the usual quasiequilibrium dynamical
equations for the moments, we should solve nonlinear (transcendent) equations. Sometimes it happens
that these equations can be written down in the explicit form (Vlasov equation, Euler equation, ten
moments Gaussian approximation in gas kinetics \cite{Ko,LPi}), but usually these equations remain in
implicit form with right-hand sides derived by a system of transcendent equations.

The post-processing is necessary for accuracy estimation. It gives us the cheapest way to improve the
solution obtained by the Legendre Integrators.

Termodynamic projector allows to transform almost arbitrary anzatz into a physically consistent
dynamic model with persistence of dissipation. The simplest example, discussed in details, is the two
peaks model for Gaussian manifold instability in polymer dynamics. This type of models opens a way to
create the computational models for the ``molecular individualism" \cite{DeGenne,Chu,LeHa1999}.

The simplest model of the molecular individualism is the ``Gaussian parallelepiped". The distribution
function is represented as a sum of $2^m$ Gaussian peaks located in the vertixes of centrally
symmetrical parallelepiped:

\begin{eqnarray*}
\Psi(q)={1 \over 2^m(2\pi)^{n/2}\sqrt{\det \Sigma}} \sum_{\varepsilon_i=\pm 1, \, (i=1, \ldots, m)}
\exp\left(-\frac{1}{2}\left(\Sigma^{-1}\left(q+\sum_{i=1}^m \varepsilon_i \varsigma_i \right), \:
q+\sum_{i=1}^m \varepsilon_i \varsigma_i\right)\right),
\end{eqnarray*}
where $n$ is dimension of configuration space, $2\varsigma_i$ is the vector of the $i$th edge of the
parallelepiped, $\Sigma$ is the one peak covariance matrix (in this model $\Sigma$ is the same for
all peaks). The macroscopic variables for this model are:
\begin{enumerate}
\item The covariance matrix $\Sigma$;
\item The set of vectors $\varsigma_i$ (or the parallelepiped edges).
\end{enumerate}
The dimension is $n(n+1)/2+mn$.

The number $m \: (m\leq n) $ is the estimated number of nonstable directions of motion (dimension of
instability). To include the nongaussian equilibrium the ``Gaussian parallelepiped" should be
deformed to nongaussian ``peaks parallelepiped". Technical details will be discussed in the separate
paper. The structure of ``peaks parallelepiped" leads to the molecular individualism in such a way:
each individual molecule belongs to a domain of a peak in configuration space. The number of these
peaks grows significantly with the dimension of instability, as $2^m$, and even if $m=3$, than the
number of peaks is 8, and one should discover 8 distinguished sorts of molecular configurations. On
the other hand, in projection on a line this amount of peaks can form a distribution without a sign
about peak structure, hence, the study of properties of ensembles (viscosity, stress coefficient,
etc.) can be without any hint to a cluster structure in configuration space.


\begin{thebibliography}{99}

\bibitem{Gibb} Gibbs, G.\ W. Elementary Principles of Statistical Mechanics, Dover, 1960.

\bibitem{Janes1}Jaynes \ E.\ T., Information theory and statistical mechanics, in:
 Statistical Physics. Brandeis Lectures, V.\ 3, 160--185 (1963).

\bibitem{Zubarev} Zubarev, D., Morozov, V.,  R\"opke, G. Statistical Mechanics of
Nonequilibrium Processes, V.\ 1, Basic Concepts, Kinetic Theory (Akademie Verlag, Berlin, 1996), V.\
2, Relaxation and Hydrodynamic Processes (Akademie Verlag, Berlin, 1997).

\bibitem{Grad} Grad, H. On the kinetic theory of rarefied gases, Comm.\ Pure and Appl.\
Math. {\bf 2(4)} (1949), 331--407.

\bibitem{KoRoz}Kogan,  A.\ M., Rozonoer, L.\ I. On the macroscopic description of kinetic processes, Dokl.
AN SSSR {\bf 158} (3) (1964), 566--569.

\bibitem{Ko} Kogan, A.\ M. Derivation of Grad--type equations and study of their properties by the method of
entropy maximization, Prikl.\ Math.\ Mech. {\bf 29} (1) (1965), 122--133.

\bibitem{Roz}Rozonoer, L.\ I. Thermodynamics of nonequilibrium processes far from equilibrium, in:
Thermodynamics and Kinetics of Biological Processes (Nauka, Moscow, 1980), 169--186.

\bibitem{G1}Gorban, A.\ N., Equilibrium Encircling. Equations of Chemical Kinetics and Their
Thermodynamic Analysis, Nauka, Novosibirsk, 1984.

\bibitem{Kark}  Karkheck, J., Stell, G., Maximization of entropy, kinetic equations, and irreversible thermodynamics
Phys.\ Rev. A {\bf 25}, 6 (1984), 3302--3327.

\bibitem{BGKTMF}Bugaenko, N. N., Gorban, A. N., Karlin, I. V. Universal Expansion of
the Triplet Distribution Function, Teoreticheskaya i Matematicheskaya Fisika, {\bf 88}, 3 (1991),
430--441 (Transl.:  Theoret. Math. Phys. (1992) 977--985).

\bibitem{Bal}Balian, R., Alhassid, Y.  and Reinhardt, H. Dissipation in many--body systems:
A geometric approach based on information theory,  Phys. Reports {\bf 131}, 1 (1986), 1--146.

\bibitem{KTGOePhA2003}Karlin, I. V., Tatarinova, L. L., Gorban, A. N., \"{O}ttinger, H. C., Irreversibility
in the short memory approximation, Physica A, to appear. Preprint online:
http://arXiv.org/abs/cond-mat/0305419 v1 18 May 2003.

\bibitem{Plenka}Gorban, A. N., Karlin, I. V., Geometry of irreversibility: The film of nonequilibrium
states, Ann. Phys., Leipzig, (2003), submitted. Preprint on-line:
http://arXiv.org/abs/cond-mat/0308331 v1 17 Aug 2003.

\bibitem{IKOePhA02}Ilg, P., Karlin, I. V., {\"O}ttinger H. C., Canonical distribution functions in polymer dynamics: I.
Dilute solutions of flexible polymers, Physica A, {\bf 315} (2002), 367--385.

\bibitem{IKOePhA03}Ilg, P., Karlin, I. V.,  Kr\"oger, M. , {\"O}ttinger H. C., Canonical distribution functions in polymer
dynamics: II Liquid--crystalline polymers, Physica A, {\bf 319} (2003), 134--150.

\bibitem{GKIOeNONNEWT2001} Gorban, A.\ N., Karlin, \ I.\ V., Ilg, P.\, and \"{O}ttinger, H.\ C.,
Corrections and enhancements of quasi--equilibrium states, J.\ Non--Newtonian Fluid Mech. {\bf 96}
(2001), 203--219.

\bibitem{GK1}Gorban, A.\ N.,  Karlin, I.\ V., Thermodynamic parameterization,
{\it Physica A}, {\bf 190} (1992), 393--404.

\bibitem{InChLANL} Gorban, A.\ N., Karlin, I.\ V., Method of invariant manifold for chemical
kinetics, Chem. Eng. Sci., to appear. Preprint online: http://arxiv.org/abs/cond-mat/0207231, 9 Jul
2002.

\bibitem{ENTR1}Gorban, A. N.,  Karlin, I. V., Family of additive entropy functions out of
thermodynamic limit, Phys. Rev. E, {\bf 67} (2003), 016104. Online:
http://arxiv.org/abs/cond-mat/0205511

\bibitem{ENTR2}Gorban, A. N.,  Karlin, I. V., \"Ottinger H. C., The additive generalization of the Boltzmann entropy,
Phys. Rev. E, {\bf 67}, 067104 (2003). Online: http://arxiv.org/abs/cond-mat/0209319

\bibitem{ENTR3}Gorban, P., Monotonically equivalent entropies and solution of additivity equation, arxiv:cond-mat/0304131,
Physica A, (2003), to appear. Online:  http://arxiv.org/pdf/cond-mat/0304131

\bibitem{Abe}Abe, S.,  Okamoto, Y. (Eds.), Nonextensive statistical mechanics and its
applications, Springer, Heidelberg, 2001.

\bibitem{Cont}Allgower E. L., Georg, K., Numerical Continuation Methods, Springer, Berlin, 1990.

\bibitem{Lin}Demmel J.W., Applied Numerical Linear Algebra, SIAM, Philadelphia, 1997.



\bibitem{MaTiWyPP}Margolin, L. G.,  Titi, E. S., Wynne, S. The postprocessing Galerkin and nonlinear Galerkin  methods
- a truncation analysis point of view,  SIAM, Journal of Numerical Analysis, (to appear)

\bibitem{Gustafsson} Gustafsson, K., Control theoretic techniques for stepsize selection in implicit Runge-Kutta methods, {ACM} Transactions on Mathematical Software, vol.20 (1994) 496--517,

\bibitem{Hairer2} Hairer, E., Norsett, S. P., Wanner, G.,
Solving ordinary differential equations I Nonstiff problems, Springer Series in Computational
Mathematics, Vol. 8, (1987) Springer-Verlag.

\bibitem{Hairer} Hairer, E., Wanner, G., Solving ordinary differential equations II Stiff and
differential-algebraic problems, Springer Series in Computational Mathematics, Vol. 14. (1991)
Springer-Verlag.

\bibitem{IK00}Ilg P.,  Karlin, I.\ V., Validity of macroscopic description in dilute polymeric solutions,
 Phys.\ Rev.\ E {\bf 62} (2000), 1441--1443.

\bibitem{FENEP} Bird, R. B.,  Dotson, R. B., Jonson, N. J., Polymer solution rheology based on a finitely extensible
bead--spring chain model,  J. Non--Newtonian Fluid Mech. {\bf 7}   (1980), 213--235; Corrigendum {\bf
8} (1981), 193,

\bibitem{IKOe99}Ilg, P.,  Karlin, I. V., \"{O}ttinger, H. C., Generating moment equations in the Doi model of
liquid--crystalline polymers, Phys. Rev. E {\bf 60} (1999), 5783--5787.

\bibitem{PGA851}Phan--Thien, N., Goh, C. G.,  Atkinson, J.\ D.,
The motion of a dumbbell molecule in a torsional flow is unstable at high Weissenberg number, J.
Non--Newtonian Fluid Mech. {\bf 18},  1 (1985), 1--17.

\bibitem{PGA852}Goh, C. G., Phan--Thien, N.,  Atkinson, J. D.,
On the stability of a dumbbell molecule in a continuous squeezing flow, Journal of Non--Newtonian
Fluid Mechanics, {\bf 18},  1 (1985), 19--23.

\bibitem{HCO} \"{O}ttinger, H.\ C., Stochastic Processes in Polymeric Fluids, Springer, Berlin, 1996.

\bibitem{Keu98}Lielens, G., Halin, P., Jaumin, I. Keunings, R. Legat, V., New closure approximations
for the kinetic theory of finitely extensible dumbbells, J.\ Non--Newtonian Fluid Mech. {\bf 76}
(1998), 249--279.

\bibitem{Martin}Kr\"oger, M., Simple models for complex nonequilibrium fluids, Phys. Reports, submitted
(2003).\\ Online: http://owl.ethz.ch/d-werk/oettinger/MK$\_$DIR/pmk80/paper.ps.

\bibitem{Herrchen}Herrchen M.,   \"{O}ttinger, H.C., A detailed comparison of various FENE dumbbell models,  J.
Non--Newtonian Fluid Mech.  {\bf 68}  (1997), 17.


\bibitem{GKTTSP94} Gorban, A.\ N., Karlin, I.\ V., Method of invariant manifolds and
regularization of acoustic spectra. {\it Transport Theory and Stat.\ Phys.}, {\bf 23} (1994),
559--632.

\bibitem{ZKD2000} Zmievskii, V.\ B., Karlin, I.\ V., Deville, M., The universal limit in
dynamics of dilute polymeric solutions. {\it Physica A}, {\bf 275(1--2)} (2000), 152--177.

\bibitem{EIT}Jou, D., Casas-V\'azquez, J., Lebon, G.,  Extended
Irreversible Thermodynamics, Springer, Berlin, 1993.

\bibitem{GKPRE96} Gorban, A. N., Karlin, I.\ V., Scattering rates versus moments: Alternative Grad equations,
Phys.\ Rev. E {\bf 54} (1996), R3109.


\bibitem{Bird}Bird,  R.\ B.,  Curtiss, C.\ F.,  Armstrong, R.\ C.,
Hassager, O., Dynamics of Polymer Liquids, 2nd edn., Wiley, New York, 1987.

\bibitem{MaKaHa2003}H\"{u}tter, M., Karlin, I. V., \"{O}ttinger, H. C., Dynamical mean-field models
from a nonequilibrium thermodynamics perspective, Phys. Rev. E, {\bf 68} (2003), 016115.

\bibitem{ArVarGZ1995-1998} Arnold, V.\ I., Varchenko, A.\ N., Gussein-Zade, S.\ M., Singularities of
differentiable maps, Boston [etc.] Brickh\"{a}user, 1985-1988. 2 vol.

\bibitem{LPi}Lifshitz E.M. and Pitaevskii L.P., Physical kinetics (Landau L.D. and Lifshitz E.M.
Course of Theoretical Physics, V.10), Pergamon Press, Oxford, 1968.

\bibitem{DeGenne}De Gennes, P. G., Molecular Individualism, Science, {\bf 276}, 5321  (1997), 1999-2000.

\bibitem{Chu}Perkins, Th. T., Smith, D. E.,  Chu, St.,  Single Polymer Dynamics in an Elongational Flow, Science,
{\bf 276}, 5321 (1997), 2016-2021.

\bibitem{LeHa1999}Leduc, P., Haber, Ch. Bao, G., Wirtz, D. Dynamics of individual flexible polymers in a shear flow,
Nature, {\bf 399} (1999), 564 - 566.


\end{thebibliography}
\end{document}